\documentclass[twocolumn]{aastex62}

\usepackage{amsthm}
\usepackage{amsmath}
\usepackage{hyperref}
\usepackage{booktabs}
\usepackage{float}
\usepackage{url}
\usepackage{natbib}

\newcommand{\Rom}{\textsc{Romulus25}}
\newcommand{\HI}{\hbox{\rmfamily H\,{\scshape i}\;}}
\newcommand\numBH{205}
\newcommand\numNoBH{197}
\newcommand\numUnder{52}
\newcommand\numMedian{99}
\newcommand\numOver{54}

\newcommand\numBHAll{306}
\newcommand\numNoBHAll{267}
\newcommand\numUnderAll{78}
\newcommand\numMedianAll{150}
\newcommand\numOverAll{78}

\begin{abstract}
	We investigate the effects of massive black hole growth on the structural evolution of dwarf galaxies within the \Rom{} cosmological hydrodynamical simulation.
	We study a sample of \numBH{} central, isolated dwarf galaxies with stellar masses $M\textsubscript{star} < 10^{10} M_\odot$ and a central BH.
    We find that the local $M\textsubscript{BH} - M\textsubscript{star}$ relation exhibits a high degree of scatter below $M\textsubscript{star} < 10^{10} M_\odot$, which we use to classify BHs as overmassive or undermassive relative to their host $M\textsubscript{star}$.
	Within isolated dwarf galaxies, only $8\%$ of undermassive BHs ever undergo a BH merger while $95\%$ of overmassive BHs grow through a mixture of BH mergers and accretion.
    We find that isolated dwarf galaxies that host overmassive BHs also follow different evolutionary tracks relative to their undermassive BH counterparts, building up their stars and dark matter earlier and experiencing star formation suppression starting around $z=2$.
	By $z=0.05$, overmassive BH hosts above $M_{star} > 10^{9} M_\odot$ are more likely to exhibit lower central stellar mass density, lower \HI gas content, and lower star formation rates than their undermassive BH counterparts.
	Our results suggest that overmassive BHs in isolated galaxies above $M\textsubscript{star} > 10^{9} M_\odot$ are capable of driving feedback, in many cases suppressing and even quenching star formation by late times.
\end{abstract}

\begin{document}

\shortauthors{Sharma et al.}
\title{Black Hole Growth and Feedback in Isolated \Rom{} Dwarf Galaxies}
\author{Ray Sharma}
\affiliation{Department of Physics and Astronomy, Rutgers, The State University of New Jersey, 136 Frelinghuysen Road, Piscataway, NJ 08854, USA}

\author{Alyson Brooks}
\affiliation{Department of Physics and Astronomy, Rutgers, The State University of New Jersey, 136 Frelinghuysen Road, Piscataway, NJ 08854, USA}

\author{Rachel S. Somerville}
\affiliation{Simons Center for Computational Astrophysics, New York, NY 10010, USA}

\author{Michael Tremmel}
\affiliation{Department of Astronomy, Yale University, 52 Hillhouse Ave, New Haven, CT 06511, USA}

\author{Jillian Bellovary}
\affiliation{Department of Physics, Queensborough Community College, City University of New York, 222--05 56th Ave, Bayside, NY 11364, USA}
\affiliation{Department of Astrophysics, American Museum of Natural History, Central Park West at 79th Street, New York, NY 10024, USA}
\affiliation{The Graduate Center, City University of New York, 365 5th Ave, New York, NY 10016, USA}

\author{Anna Wright}
\affiliation{Department of Physics and Astronomy, Rutgers, The State University of New Jersey, 136 Frelinghuysen Road, Piscataway, NJ 08854, USA}

\author{Thomas Quinn}
\affiliation{Department of Astronomy, University of Washington, PO Box 351580, Seattle, WA 98195, USA}

\section{Introduction} \label{sec:introduction}

	Evidence has built over the last two decades demonstrating that massive black holes (BHs) are ubiquitous in massive galaxies~\citep{Kormendy1995,Kormendy2013} though less common in low-mass galaxies~\citep{Shields2008,Reines2013,Moran2014}.
	BHs in massive galaxies are thought to co-evolve with their host through a number of processes connecting BH growth to host growth~\citep{Fabian2012,Somerville2015,Heckman2014}.
	There are numerous detections of active galactic nuclei (AGN) within dwarf galaxies~\citep{Reines2013,Reines2015,Lemons2015,Pardo2016,Baldassare2017,Ahn2018,Baldassare2018,Martin-Navarro2018,Baldassare2020}.
	Recent studies have also estimated dynamical masses for many weakly accreting BHs within dwarf galaxies~\citep{Reines2015,Nguyen2017,Nguyen2018,Nguyen2019,Nguyen2020}.
	However, studying the role that BHs play in the evolution of dwarf galaxies has only become possible in recent years.

	Massive BHs have been observed to have a significant impact on gas and stars within dwarf galaxies.
	\citet{Penny2018} identify $6$ dwarf galaxies $\left(M\textsubscript{star} < 5 \times 10^{9} M_\odot\right)$ within SDSS that exhibit 1) kinematically offset stars and ionized gas components, and 2) strongly AGN-like emission line ratios identifying the AGN as the primary source of gas ionization.
	\citet{Manzano-King2019} identify $9$ SDSS dwarf galaxies with AGN-like narrow line emission and indirect evidence of AGN-driven star formation suppression.
	\citet{Bradford2018} find evidence of reduced \HI gas mass in isolated dwarf galaxies $\left(10^{9.2} < M\textsubscript{star} < 10^{9.5} M_\odot\right)$ exhibiting AGN-like ionizing radiation, using \HI-selected data from ALFALFA and optically-selected data from SDSS\@.
	Similarly,~\citet{Dickey2019} find a connection between AGN-like ionizing radiation and quenching of star formation in the host galaxy, using Keck/ESI spectra of isolated dwarf galaxies $\left(10^{9} < M\textsubscript{star} < 10^{9.5} M_\odot\right)$.
	\citet{Silk2017} find, within the dense progenitors of modern dwarf galaxies, BHs can generate an order of magnitude more power than supernovae (SNe).
	Silk argues that early BH feedback may play a role in explaining a number of problems in modelling structure formation, such as suppression of luminous dwarf formation~\citep{Trujillo-Gomez2015}, the missing cold baryons in the local universe~\citep{Bregman2015}, and the source of ionizing photons during reionization~\citep{Madau2015}.

	Although observations indicate BHs may impact dwarf galaxy evolution, detecting BHs and disentangling the precise role of BHs in dwarf galaxies is difficult for a few reasons:
	First, the sample sizes involved in surveys of BHs in dwarf galaxies may be restricted by the fraction of galaxies that host BHs~\citep{Greene2012}.
	We define the BH occupation fraction to be the total number of BH-hosting galaxies compared to the total number of galaxies, for a given stellar mass.
	Recent work finds the BH occupation fraction is sensitive to the BH formation mechanism~\citep{Ricarte2018}, and may drop to zero for sufficiently low-mass galaxies.
	Observations are consistent with a roughly $100\%$ occupation fraction down to $M\textsubscript{star} \sim 10^{9} M_\odot$~\citep{Miller2015,Nguyen2018,Nguyen2020,Baldassare2020}.
	Simulations suggest the BH occupation fraction plummets below stellar mass $M\textsubscript{star} < 10^{9} M_\odot$~\citep{Habouzit2017,Ricarte2018,Bellovary2019}.
	Measuring the true BH occupation fraction is further restricted by the difficulties in detecting weakly accreting BHs outside of the local universe~\citep{Bellovary2019}.

	Second, detection of BH-induced star formation suppression is impeded by evidence that low-mass galaxy quenching in group environments is closely tied to tidal effects~\citep{Penny2016}.
	Over $99 \%$ of quenched dwarf galaxies seem to be found within $1.5$ Mpc of a galaxy with Milky Way mass or greater~\citep{Geha2012}.
	Although AGN may be more easily observed in group environments than in the field~\citep{Penny2018}, distinguishing the dominant quenching source within groups proves challenging.

	Third, BH detection surveys often suffer from dust obscuration as well as sample contamination from various sources.
	A large portion of AGN in dwarf galaxies may be completely missed in optical or x-ray surveys due to heavy dust obscuration~\citep{Chen2017,Liu2018}.
 	X-ray binaries can further contaminate surveys that identify AGN through x-ray detection~\citep{Yuan2014,Miller2015}.
	The effects of star formation and BH accretion on ionizing the interstellar medium are difficult (if not impossible) to disentangle in low-metallicity galaxies~\citep{Groves2006,Kewley2013}.
	\citet{Reines2013} find that identifying AGN solely through ionizing radiation diagnostics can lead to missed detections.

	The difficulties in detecting BHs and disentangling their impact necessitate the use of simulations to further study them.
	High-resolution cosmological simulations provide one of the best laboratories for predicting BH population growth and behavior, over a wide range of host stellar masses and redshifts.
	In the past few years, simulations have successfully begun to reach the resolutions required to capture BH physics within low-mass galaxies.
	\citet{Habouzit2017} find that SNe feedback within simulated low-mass \textsc{SuperChunky} galaxies $\left(M\textsubscript{halo} \lesssim 10^{10.5} M_\odot\right)$ can suppress accretion onto the central BH.
	By modelling AGN and SN outflows~\citet{Dashyan2018} suggest extended periods of BH growth and activity can drive gas out of dwarf galaxies more efficiently than SNe.
	Cosmological simulations run by~\citet{Barai2019} find BHs are capable of quenching dwarf galaxies through BH feedback by $z=4$.
	Using high-resolution, cosmological, zoom-in simulations, \citet{Bellovary2019} constrain the cosmic BH occupation fraction in low-mass galaxies, and find BHs within dwarf galaxies grow little throughout their lifetime.
	~\citet{Koudmani2019} test various models of AGN feedback in simulated dwarf galaxies, finding that AGN can significantly enhance outflow temperatures and velocities from stellar processes, inhibiting gas inflows.

	Simulations that include BHs face a number of numerical obstacles.
	Simulating BH growth and feedback requires simulating a large range of scales with high resolution.
	To mitigate these limitations, simulations incorporate sub-grid prescriptions designed to model physics occurring below the resolution limits.
	Sub-grid prescriptions have been shown to reproduce observed properties of BHs and their hosts~\citep{Hirschmann2014a,Sijacki2014,Volonteri2016,Habouzit2017,Habouzit2019}.
	However, the chosen prescriptions for BH formation, dynamics, and accretion can drastically impact the predicted BH growth, occupation fraction, and assembly history of the galaxy \citep{Ricarte2018}.
	Furthermore, the especially high resolutions required to capture BH physics in dwarf galaxies force most cosmological simulations to set a halo mass treshold for BH formation.

	In this paper we analyze results from the high-resolution \Rom{} cosmological simulation to understand how BHs in dwarf galaxies grow and interact with their environments.
	We explore the growth mechanisms and evolutionary history of BHs, the connection between BH growth and host galaxy growth, and the possibility of significant BH feedback in \Rom{} dwarf galaxies.
	\Rom{} is well-suited for this analysis since it is currently the only cosmological simulation capable of forming BHs in low-mass galaxies at high enough resolution and with sufficiently detailed physical prescriptions to accurately track their dynamics and growth (see \S\ref{subsec:black-hole-dynamics} for details).
	To model the evolution of BHs, simulations must be able to accurately account for dynamical friction, which requires dark matter and stars at higher mass resolution than BHs to realistically model BH dynamics \citep{Tremmel2015}. Hence the BH seed mass in \Rom{} is closely tied to the simulation resolution.
	\Rom{} achieves BHs that are $\sim 3$ times more massive than dark matter particles, while also being one of the highest resolution simulation volumes to date.
	Additionally, while most simulations set a halo mass threshold for BH formation, \Rom{} does not have a priori assumption of the BH occupation fraction.
	Our analysis also gives insight into how the BH physics and sub-grid recipes within \Rom{} affect BH growth in dwarf galaxies.

	In Section~\ref{sec:simulation} we describe the physics of the \Rom{} cosmological simulation.
	In Section~\ref{sec:results} we explore the connection between BH growth and properties of the host dwarf galaxy.
	In particular, we explore how dwarf galaxies can form significantly overmassive and undermassive BHs, and how such BHs may drive evolutionary differences between their hosts.
	We discuss the consequences of scatter in the $M\textsubscript{BH}-M\textsubscript{star}$ relation as well as the impact of BHs on star and gas properties in dwarf galaxies.

\section{\textsc{Romulus} Simulation Suite} \label{sec:simulation}

	\subsection{Simulation Properties} \label{subsec:simulation-properties}
		The \textsc{Romulus} suite of cosmological simulations were run using the Tree + SPH code ChaNGa~\citep{Menon2015} which inherits baryonic prescriptions from \textsc{Gasoline} and \textsc{Gasoline2}~\citep{Wadsley2004,Wadsley2008,Stinson2006,Shen2010,Wadsley2017}.
		In this work we analyze \Rom{}, the 25 Mpc per side uniform box with periodic boundary conditions.
		We analyze \Rom{} because of its large, uniform sample of low-mass galaxies with BHs run to $z = 0$.

		\Rom{} has comparable mass resolution to recent cosmological simulations such as \textsc{IllustrisTNG}~\citep{Springel2018} and Horizon-AGN~\citep{Volonteri2016}, as well as force resolution comparable to the highest resolution runs of EAGLE~\citep{Schaye2015}.
		\Rom{} resolves gravity with a Plummer equivalent force softening of $\epsilon\textsubscript{g} = 250$ pc.
		Typically, the number of gas particles in a simulation is equal to the number of dark matter particles, while the relative masses are set according to the cosmic baryon fraction.
		\Rom{} instead contains $3.375\times$ more dark matter particles than gas particles, with dark matter particles of mass $3.39 \times 10^5 M_\odot$ and gas particles of mass $2.12 \times 10^5 M_\odot$.
		This ``oversampling'' of dark matter provides \Rom{} with better resolved BH dynamics~\citep{Tremmel2015}.
		The \textsc{Romulus} suite of simulations were run with a Planck 2014 $\Lambda CDM$ cosmology, with $\Omega\textsubscript{m} = 0.3086$, $\Omega_\Lambda = 0.6914$, $h = 0.6777$, and $\sigma\textsubscript{8} = 0.8288$~\citep{PlanckCollaboration2014}.

		Halos were identified using the \textsc{Amiga} Halo Finder~\citep{Knollmann2009} and analyzed using the simulation analysis code \textsc{Pynbody}~\citep{Pontzen2013}.
		Key properties were organized into a \textsc{Tangos} database~\citep{Pontzen2018}.

		We calculate halo mass, $M\textsubscript{200}$, such that:
			\begin{align}
				M\textsubscript{200} = 4\pi \rho_c \Delta\textsubscript{h} R_{200}^3,
			\end{align}
		where $\rho_c$ is the critical density, $\Delta\textsubscript{h} = 200$ is the overdensity threshold, and $R\textsubscript{200}$ is the halo radius.

		Star formation in \Rom{} is regulated by the star formation efficiency, the efficiency of supernova energy injection into the interstellar medium, and the density/temperature threshold beyond which stars are allowed to form.
		SN feedback follows the blastwave prescription from~\citet{Stinson2006}.

        Parameters governing star formation were constrained based on a series of 80 'zoom-in' \citep{Governato2009} simulations of 4 galaxies with halo masses $10^{10.5} - 10^{12} M_\odot$, run without BH physics, with the aim of reproducing a set of observed $z=0$ scaling relations for galaxies.
        The parameter spaces were explored using the Kriging algorithm and graded by agreement with the stellar mass - halo mass relation~\citep{Moster2013}, the relationship between stellar mass, angular momentum, and morphology ~\citep{Obreschkow2014}, and the $\HI$ gas - stellar mass relation~\citep{Cannon2011,Haynes2011}.
		The Kriging algorithm efficiently traverses the parameter space and penalizes parameter selections that lead to deviations from the observed scaling relations.
		The parameters governing BH growth were afterward constrained in a similar fashion  (see Section \S\ref{subsec:black-hole-accretion-and-feedback}).
		
		\Rom{} adopts:
			\begin{itemize}
				\item SF efficiency, $c_\ast= 0.15$
				\item Gas density treshold, $n_\ast = 0.2$m$\textsubscript{p}$ cm$^{-3}$
				\item SNe coupling efficiency, $\epsilon\textsubscript{SN} = 0.75$
			\end{itemize}

		\Rom{} incorporates prescriptions for metal and thermal diffusion from~\citet{Shen2010,Governato2015}, low-temperature radiative cooling as in~\citet{Guedes2011}.
		The SNe feedback is based on a ~\citet{Kroupa2001} initial mass function.
		Specifics of the physics and calibration process for \Rom{} are detailed in~\citet{Tremmel2017}.

	\subsection{Black Hole Seeding} \label{subsec:black-hole-seeding}

		Simulations commonly seed BHs by choosing a halo mass threshold above which galaxies are allowed to form a BH~\citep{Sijacki2014,Angles-Alcazar2017}.
		BHs are then formed from star-forming gas and placed in the halo center~\citep{Agarwal2014}.
		Instead, \Rom{} seeds BHs based on conditions of the pre-collapse gas using characteristics of direct-collapse seeding~\citep{Haiman2013}.
		In this work we qualitatively define massive BHs as BHs with masses relevant to direct-collapse seeding models $\left(M\textsubscript{BH} \gtrsim 10^{5} M_\odot\right)$.
		\Rom{} seeds massive BHs at high redshift $\left(z \gtrsim 5\right)$ without assumptions of the BH occupation fraction.
		A gas particle is marked to form a BH if it has:
			\begin{itemize}
			  \item Low metallicity, Z $< 3\times 10^{-4}$
			  \item High density, $15\times$ higher than the SF threshold
			  \item Temperature between $9500 - 10000$K
			\end{itemize}
		In other words a gas particle will form a BH if the gas is set to collapse quickly and cool slowly, following predicted sites of direct-collapse seeding~\citep{Begelman2006,Johnson2012,Volonteri2012,Haiman2013,Reines2016}.
		In particular, a low metallicity threshold prevents premature gas fragmentation and pushes seed formation to early times when gas has undergone little metal enrichment~\citep{Greene2012}.
		The values were chosen to restrict BH growth to the highest density regions in the early universe, where BHs can undergo rapid accretion.
		Choosing lower metallicity or colder temperature thresholds was found to bias formation away from the densest regions~\citep{Tremmel2017}.

		Once these gas conditions are met a BH is seeded at a mass of $M\textsubscript{BH} = 10^{6} M_\odot$.
		Seeding accretes mass from nearby gas particles to conserve total mass and simulate rapid, unresolved growth thought to exceed $0.1 M_\odot$ yr$^{-1}$~\citep{Hosokawa2013,Schleicher2013}.
		This seed mass is high relative to some other simulations, such as \textsc{IllustrisTNG}~\citep{Nelson2019}.
		Dynamical BH estimates (e.g., \citet{Nguyen2020}) also indicate that BHs in dwarf galaxies can fall to masses well below $10^{6} M_\odot$~.
		However, seeding BHs at a higher mass ensures that both BH dynamics and gas accretion are well-resolved throughout their lifetimes~\citep{Tremmel2015}.
		Further, the formation mechanism ensures BHs are only seeded in regions with dense, collapsing gas that is unlikely to form stars, and hence are more likely to grow BHs rapidly.
		For more detailed caveats related to the BH seeding mechanism, see Section \S\ref{subsec:caveats}.

	\subsection{Black Hole Dynamics}\label{subsec:black-hole-dynamics}

		Dynamical friction between BHs and their hosts causes BHs to sink toward the galaxy center~\citep{Kazantzidis2005,Pfister2017}.
		Dynamical friction interactions occur at both large scales~\citep{Colpi1999} and scales below the resolution of most cosmological simulations.
		Common practice in cosmological simulations is to reposition the BH along local potential gradients as it begins to migrate, forcefully re-centering it.
		However, this method suppresses BH motion around the galaxy and artificially inflates BH growth rates~\citep{Tremmel2017}.
		\Rom{} instead incorporates a dynamical friction sub-grid recipe shown to reproduce realistic BH sinking timescales.
		This recipe allows BHs to naturally ``wander'' within galaxies~\citep{Tremmel2018,Bellovary2019,Reines2020}.
		\citet{Tremmel2015} find the spatial resolution in \Rom{} combined with its oversampling of dark matter also avoids unrealistic numerical heating of BHs found in lower resolution simulations.

		Assuming an isotropic velocity distribution of particles within a gravitational softening length, $\epsilon\textsubscript{g}$, from the BH\@, we can write the acceleration due to dynamical friction~\citep{Chandrasekhar1943}:
			\begin{align}
				\textbf{a}\textsubscript{DF} = -4\pi G^2 M\textsubscript{BH} \rho(< v\textsubscript{BH})\ln \Lambda \frac{\textbf{v}\textsubscript{BH}}{v^3\textsubscript{BH}},
			\end{align}
		where $M\textsubscript{BH}$ is the mass of the BH, $\rho( < v\textsubscript{BH})$ is the density of nearby particles moving slower than the BH, $\ln\Lambda$ is the Coulomb logarithm, and \textbf{v}$_{BH}$ is the velocity of the BH\@.
		Velocities are calculated relative to the local center of mass within the smoothing kernel.
		The Coulomb logarithm is approximated by $\ln\Lambda \sim \ln \frac{b_{\max}}{b_{\min}}$, where $b_{\max}$ and $b_{\min}$ are respectively the maximum and minimum impact parameters of the surrounding particles.
		The maximum impact parameter is restricted to $b_{\max} = \epsilon\textsubscript{g}$ to avoid double counting of resolved dynamical frictional forces.
		The minimum impact parameter is restricted to the $90^\circ$ deflection radius with a lower limit of the Schwarzschild radius of the BH\@.
		Acceleration is calculated from the nearest 64 collisionless particles.
		Mergers occur when BHs fall within $2 \epsilon\textsubscript{g}$ of one another and have low enough relative velocity to be considered gravitationally bound.

	\subsection{Black Hole Accretion and Feedback} \label{subsec:black-hole-accretion-and-feedback}

		BH accretion is handled through a Bondi-Hoyle prescription modified to incorporate angular momentum support on resolved scales.
		The accretion rate is driven by mass flux across a resolved accretion radius, defined as the radius where gravitational potential balances with the minimally resolved thermal energy of the surrounding gas.
		Accretion rates are averaged over the smallest simulation timestep at a given time, typically $10^4 - 10^5$ years.
		We can write the accretion rate depending on whether the dominant gas motion is rotational, $v_{\theta}$, or bulk flow, $v\textsubscript{bulk}$:
			\begin{align} \label{Accretion Rate}
				\dot{M}= \alpha \times
					\begin{cases}
						\frac{\pi G^2 M^2_{BH} \rho}{\left(v_{bulk}^2 + c_s^2\right)^{3/2}} &\text{if } v_{bulk} > v_\theta\\
						\frac{\pi G^2 M^2_{BH} \rho c_s}{\left(v_\theta^2 + c_s^2\right)^2} &\text{if } v_{bulk} < v_\theta,
					\end{cases}
			\end{align}
			where
			\[
				\alpha =
					\begin{cases}
					   \left(\frac{n\textsubscript{gas}}{n\textsubscript{*}}\right)^\beta &\text{if } n\textsubscript{gas} \geq n\textsubscript{*}\\
					   1 &\text{if } n\textsubscript{gas} \leq n\textsubscript{*},
					\end{cases}
			\]
		is the density-dependent boost factor that corrects for underestimates of accretion rate due to resolution limitations~\citep{Booth2009}, $\beta = 2$ is the corresponding boost coefficient, $n\textsubscript{gas}$ is the number density of the surrounding gas, $n\textsubscript{*}$ is the star formation density threshold, $\rho$ is the mass density of the surrounding gas, $c\textsubscript{s}$ is the local sound speed, $v_\theta$ is the rotational velocity of the surrounding gas at the smallest resolved scales, and $v\textsubscript{bulk}$ is the bulk velocity of the surrounding gas.
		This calculation is performed over the 32 nearest particles.
		BH accretion in \Rom{} is Eddington limited.

	    BH feedback is handled through a sub-grid recipe similar to the blastwave SN feedback model.
		Thermal energy from accretion is isotropically transferred to the nearest 32 gas particles, weighted by the SPH kernel.
		To ensure realistic dissipation of feedback energy, gas particles that receive energy are stopped from cooling for roughly the gas dynamical timestep over which the accretion is calculated~\citep{Tremmel2017}.
		The energy coupled to the surrounding gas is
			\begin{align} \label{Feedback Prescription}
				E = \epsilon\textsubscript{r} \epsilon\textsubscript{f} \dot{M} c^2 dt,
			\end{align}
	    where $\epsilon\textsubscript{r} = 0.1$ is the assumed radiative efficiency of the BH, and $\epsilon\textsubscript{f} = 0.02$ is the coupling efficiency of the thermal energy to the surrounding gas (see below).
		This form of BH feedback has been shown to efficiently quench galaxies with halo masses above a few $10^{11} M_\odot$~\citep{Pontzen2017}.

		The accretion efficiency, $\beta$, and coupling efficiency, $\epsilon\textsubscript{f}$, were constrained through 48 zoom-in simulations of the same 4 galaxies used for constraining star formation parameters, run with BH physics.
		The star formation parameters were left unchanged from the initial parameter search and instead the BH parameters were allowed to change.
		The parameter space was explored using the Kriging algorithm and graded by agreement with an empirical $z=0$ BH mass - stellar mass relation~\citep{Schramm2013}.
		The results of this parameter search were also used in the high resolution cosmological hydrodynamic, galaxy cluster simulation, \textsc{RomulusC}~\citep{Tremmel2019}.

\section{Results} \label{sec:results}

	\begin{table}[]
		\begin{tabular}{@{}lcc@{}}
			\toprule
			\multicolumn{1}{c}{Sample} & All M$\textsubscript{star}$ & $10^{8} <\,$M$\textsubscript{star} < 10^{10}\,$M$_\odot$ \\ \midrule
			Undermassive & \numUnderAll{}    & \numUnder{}                                  \\
			Median       & \numMedianAll{}   & \numMedian{}                                 \\
			Overmassive  & \numOverAll{}     & \numOver{}                                   \\
			\midrule
			With BHs     & \numBHAll{}       & \numBH{}                                     \\
			Without BHs  & \numNoBHAll{}     & \numNoBH{}                                   \\ \bottomrule
			\end{tabular}
			\caption{Sample sizes of isolated galaxies in \Rom{} with and without massive BHs, distinguishing between BH classifications. \label{sample-table}}
	\end{table}

	We restrict our sample in a few ways.
	We select galaxies with stellar masses between $10^{8} < M\textsubscript{star} < 10^{10} M_\odot$ at $z=0.05$, where the lower threshold ensures galaxies have at least several hundred star particles.
	We also restrict our sample to central, relatively isolated galaxies to better separate the effects of BHs from the effects of group environments on dwarf galaxy evolution.
	A halo is considered isolated if it is farther than one halo radius $R\textsubscript{200}$ from another halo of equal or greater mass at $z=0$.
	Following~\citet{Geha2012}, galaxies below $M\textsubscript{star} < 10^{10} M_\odot$ must be farther than $1.5$ Mpc of a galaxy with $M\textsubscript{star} > 2.5\times 10^{10} M_\odot$ to be considered isolated.
	Geha et al find evidence of environmental quenching of low-mass galaxies within such scales, although they focused on galaxies with $M\textsubscript{star} < 10^{9} M_\odot$. Since many BHs have been found in dwarfs between $10^{9} < M\textsubscript{star} < 10^{10} M_\odot$, we study the full range of \Rom{} dwarfs up to $M\textsubscript{star} < 10^{10} M_\odot$ but still apply the isolation criteria above.
	Finally, we define the central BH to be the most massive BH within $2$ kpc of the halo center, since \Rom{} halos can contain many BHs at varying radii from the center~\citep{Tremmel2018}. Once we restrict to only hosts of central BHs, the mean BH distance for our galaxy sample drops from $1$ kpc to $170$ pc.
			
	These restrictions give us a sample of \numBH{} isolated dwarf galaxies with central BHs, as well as a sample of \numNoBH{} isolated dwarf galaxies entirely without BHs. We summarize our sample sizes in Table \ref{sample-table}, alongside sample sizes for specific subsets of our data (see Section \ref{subsec:results1} for details).

	We use stellar mass corrections from~\citet{Munshi2013} that account for the impact of aperture photometry on observed stellar masses.
	These corrections allow us to perform a more ``apples-to-apples'' comparison between simulated and observed stellar masses.
	Following Munshi et al, we correct stellar masses such that $M\textsubscript{star, obs} = 0.6 M\textsubscript{star, sim}$.

	\subsection{$M\textsubscript{BH} - M\textsubscript{star}$ Relation} \label{subsec:results1}

		\begin{figure}
			\plotone{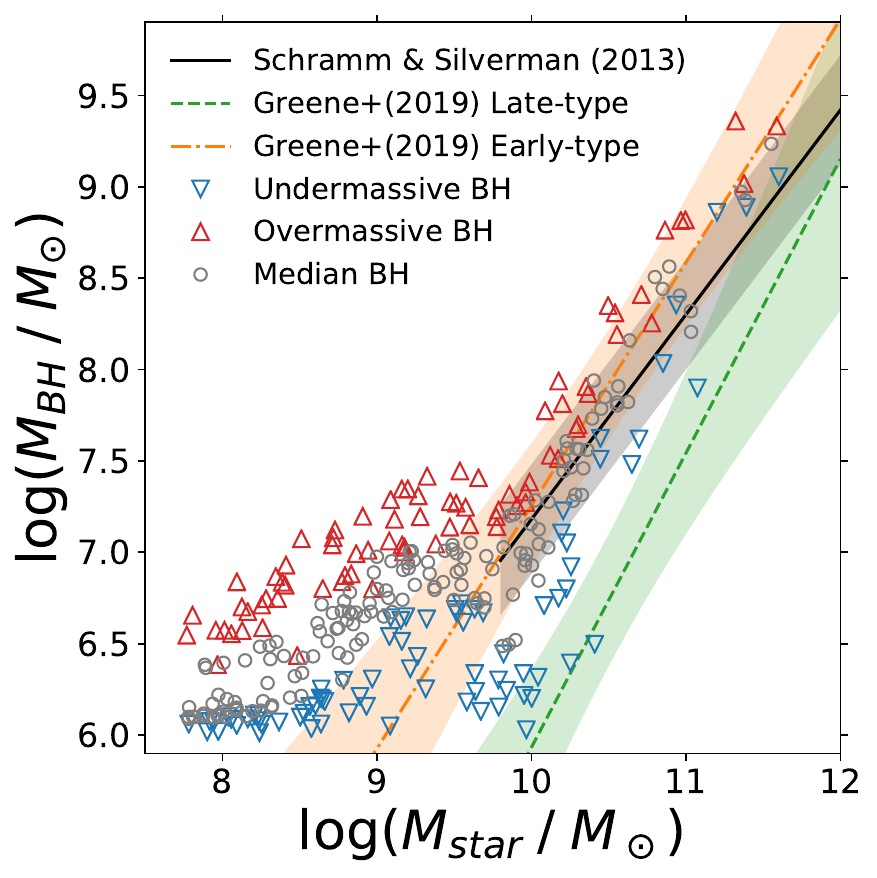}
			\caption{The $z=0.05$ BH mass versus stellar mass relation for all isolated galaxies in \Rom{} with central BHs.
				The relation has large scatter in $M\textsubscript{BH}$ below $M\textsubscript{star} < 10^{10} M_\odot$, but becomes well constrained above $M\textsubscript{star} > 10^{10} M_\odot$.
				Galaxies are colored according to whether the hosted BH is overmassive (red triangles), undermassive (blue inverted triangles), or median (grey circles) (see text for definitions.)
				We compare our relation to both the early-type (orange dash dotted) and late-type (green dashed) relations compiled by ~\citet{Greene2019}, as well as the x-ray selected broad-line AGN relation from ~\citet{Schramm2013} (black solid).
				Shaded regions indicate $1\sigma$ observational uncertainties.
				Above $M\textsubscript{star} \gtrsim 10^{10} M_\odot$, \Rom{} shows agreement with the relation from Schramm \& Silverman, as well as with the early-type relation from Greene et al.
				Below $M\textsubscript{star} < 10^{10} M_\odot$, the total BH mass is dominated by the BH seed mass and we find significant deviation away from observed relations.
				\label{Mbh-Mstar}}
		\end{figure}

		\begin{figure}
			\plotone{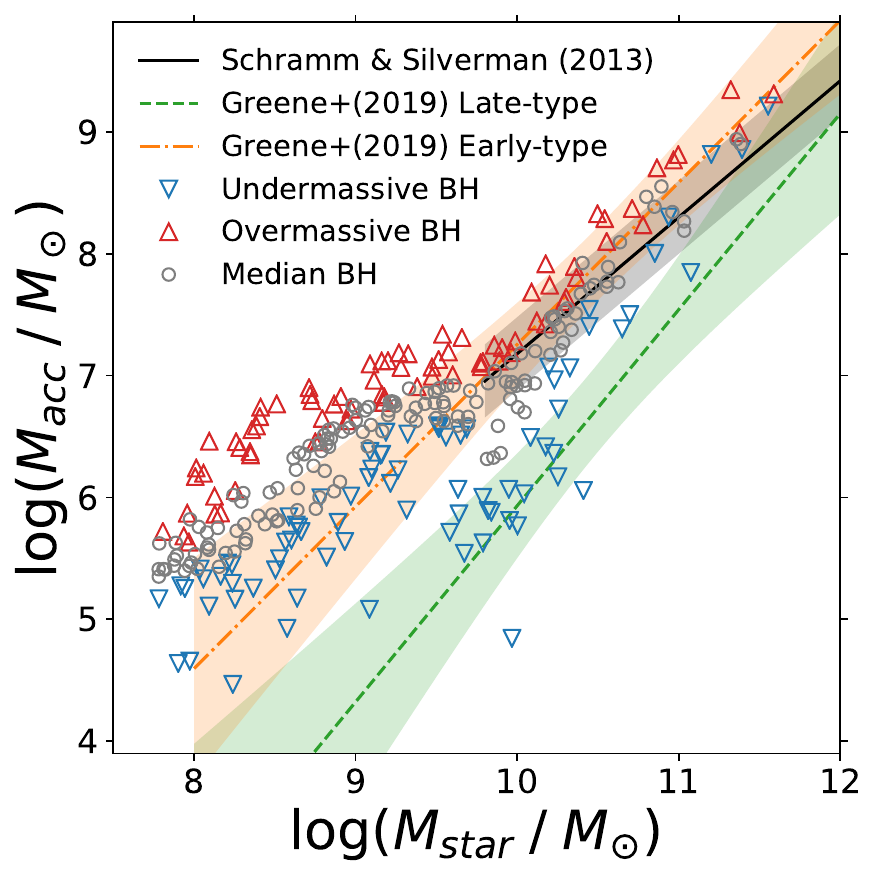}
			\caption{The $z=0.05$ total accreted BH mass versus stellar mass relation for all isolated galaxies in \Rom{} with central BHs.
				The relation has large scatter in $M\textsubscript{acc}$ below $M\textsubscript{star} < 10^{10} M_\odot$, as in Figure~\ref{Mbh-Mstar}.
				Galaxies are colored according to whether the hosted BH is overmassive (red triangles), undermassive (blue inverted triangles), or median (grey circles).
				\Rom{} shows agreement with the relation from \citet{Schramm2013}, as well as with the early-type relation from \citet{Greene2019}.
				Below $M\textsubscript{star} < 10^{10} M_\odot$, overmassive BHs tend to fall above the early-type relation.
				\label{Macc-Mstar}}
		\end{figure}

		Figure~\ref{Mbh-Mstar} shows the $z=0.05$ $M\textsubscript{BH} - M\textsubscript{star}$ relation for all isolated \Rom{} galaxies with central BHs.
		Galaxies below $M\textsubscript{star} < 10^{10} M_\odot$ exhibit a high degree of scatter in $M\textsubscript{BH}$, which is a phenomenon that has been predicted and observed before.
		Using semi-analytic models~\citet{Volonteri2009} find that the evolution of BHs toward the local $M\textsubscript{BH} - \sigma$ relation is dependent on both the BH growth history and seed BH mass.
		They find low-mass BHs exhibit a higher amount of scatter on the $M\textsubscript{BH} - \sigma$ relation than higher mass BHs.
		Using cosmological hydrodynamical simulations~\cite{Barai2019} similarly find high scatter in the $M\textsubscript{BH} - M\textsubscript{star}$ relation at low stellar masses.
		\citet{Reines2015} find that the $M\textsubscript{BH} - M\textsubscript{star}$ relation observed in high-mass galaxies breaks down in low-mass galaxies.
		They find low-mass star-forming galaxies may instead follow a  relation with lower BH masses than expected.
		In high mass galaxies,~\citet{Shankar2016} find that dynamical estimates of the local $M\textsubscript{BH} - M\textsubscript{star}$ relation are biased by the requirement that the BH sphere of influence be fully resolved.
		Shankar et al. find that corrections to this bias place dynamical $M\textsubscript{BH} - M\textsubscript{star}$ relations in closer agreement with AGN-derived relations, and eliminate much of the perceived scatter in $M\textsubscript{BH}$ at high $M\textsubscript{star}$.

		Galaxies around $M\textsubscript{star} \sim 10^{8} M_\odot$ tend to clump up at the BH seed mass and twice the seed mass.
		The simulation BH seeding mechanism introduces a floor that likely inflates BH masses in the lightest dwarf galaxies relative to observations.
		We do not remove hosts of seed mass BHs from the sample since it is unclear what additional biases would be introduced relative to a higher-resolution seeding mechanism.
		For some analyses that follow we separate the sample into two mass bins since resolution may impact results at the lowest masses.

		We compare with observed relations from~\citet[][SS13]{Schramm2013} and ~\citet{Greene2019}.
		\Rom{} agrees with the Greene et al relation for early-type galaxies at stellar masses $M\textsubscript{star} > 10^{10} M_\odot$, as well as with the SS13 relation.
		Above $M\textsubscript{star} \gtrsim 10^{10} M_\odot$, galaxies follow the SS13 relation in part because the BH physics is tuned to match with the low-mass end of the SS13 relation.
		For similar reasons, \Rom{} does not agree with the Greene et al.~relation for late-type galaxies.

		The divergence between the various relations may be due to a mixture of a few effects: morphological differences in the host galaxy, differences in accretion efficiencies between BHs, and uncertainties in observational BH mass estimators.
		\citet{Reines2015} find significant differences between their $M\textsubscript{BH} - M\textsubscript{star}$ relations for bulge-dominated galaxies and AGN.
		Their bulge-dominated galaxy sample has dynamical BH mass estimates while their AGN sample contains a mixture of pseudobulges and classical bulges with broad H$\alpha$ virial + reverberation-mapped BH masses.
		They find that the scatter in the combined $M\textsubscript{BH} - M\textsubscript{star}$ relation is closely related to the morphology of the host galaxy, where bulge-dominated galaxies sit on a relation with similar slope but higher normalization than spiral galaxies.
		SS13 do not make morphological distinctions in their sample and base their relation on x-ray selected AGN with broad MgII virial BH masses.
		Similarly,~\citet{Davis2018} find late-type galaxies follow a steeper $M\textsubscript{BH} - M\textsubscript{star}$ relation than early-type galaxies down to $M\textsubscript{star} \sim 10^{10.5} M_\odot$, which~\citet{Sahu2019} find is fundamentally a difference in bulge morphology of the host galaxies.
		
		\citet{Trump2011} find that broad emission line regions are only observed in rapidly accreting AGN with luminosities greater than $10^{-2} L\textsubscript{Edd}$, where $L\textsubscript{Edd}$ is the Eddington luminosity.
		By analyzing \Rom{}, \citet{Ricarte2019} find that high Eddington ratios are associated with BHs that exhibit systematically lower masses than expected for their host stellar mass.
		
		Indirect measurements of BH masses within AGN are typically thought to yield large uncertainties, dependent on their luminosity and redshift (e.g., \citealt{Shen2012,Kelly2013}).
        Virial BH mass estimates of AGN require assumptions of the geometry and orientation of the broad-line region~\citet{Denney2010,Barth2011}, and may be biased by non-gravitational contributions to broad-line widths~\citet{Krolik2001}.
        However, \citet{Reines2015} argue that the difference between $M\textsubscript{BH} - M\textsubscript{star}$ relations for AGN versus elliptical galaxies is far larger than can be explained by uncertainties in virial BH masses.
        They further find their virial BH masses would need an unreasonably high virial factor in order to match the elliptical galaxy relation.

		We quantify scatter in the $M\textsubscript{BH} - M\textsubscript{star}$ relation by the residual away from the median $M\textsubscript{BH}$ in bins of $M\textsubscript{star}$.
		Within each bin we classify BHs with masses in the bottom 25\% as ``undermassive'' while BHs within the top 25\% are classified as ``overmassive.''
		BHs falling between the two quartiles are classified as ``median.''
		We summarize our sample sizes in Table \ref{sample-table}.
		We show this classification scheme at all stellar masses, though the classification is most meaningful in dwarf galaxies where the central BHs show a high degree of scatter in $M\textsubscript{BH}$.
		A similar classification scheme is built by~\citet{Li2019} for $M\textsubscript{star} > 10^{10} M_\odot$ \textsc{Illustris} galaxies and $M\textsubscript{star} > 10^{9} M_\odot$ \textsc{TNG100} galaxies, though they instead define overmassive and undermassive BHs on the $M\textsubscript{BH} - \sigma$ relation.

		It should be emphasized that current observations find BHs in dwarf galaxies can have masses lower than the resolution limit of \Rom{}.
		~\citet{Baldassare2015} find a BH in the dwarf galaxy RGG 118 with a virial mass of $M_{BH} = 5 \times 10^{4} M_\odot$.
		~\citet{Nguyen2017} calculate a dynamical BH mass of $M_{BH} = 1.5 \times 10^{5} M_\odot$ in the nearby dwarf galaxy NGC 404.
		~\citet{Nguyen2020} improve dynamical BH masses for three nearby low-mass galaxies, where all three show BH masses below 1 million solar masses, and one shows a dynamical mass of $M_{BH} = 6.8 \times 10^{3} M_\odot$.
		~\citet{Graham2019} and ~\citet{Graham2019a} use BH scaling relations to predict BH masses as low as $M\textsubscript{BH} \sim 10^{4} M_\odot$ in low-mass galaxies in the Virgo cluster.

		Following~\citet{Ricarte2019}, we can partially compensate for the effects of the BH seeding mechanism and define the total mass a given BH has grown via accretion, $M\textsubscript{acc}$.
		This definition completely excludes the contributions of BH seeding and only counts accretion onto every progenitor within the BH merger tree.
		We are able to trace $M\textsubscript{acc}$ well below the BH seed mass because of the high resolution of gas accretion in \Rom{}.
		Figure~\ref{Macc-Mstar} shows the $M\textsubscript{acc} - M\textsubscript{star}$ relation for all isolated \Rom{} galaxies with central BHs, compared with observed $M\textsubscript{BH} - M\textsubscript{star}$ relations.
		We find that the $M\textsubscript{acc} - M\textsubscript{star}$ relation continues linearly below the BH seed mass, and shows better agreement with the observed relations.
		Galaxies below $M\textsubscript{star} < 10^{10} M_\odot$ still show a high degree of scatter in $M\textsubscript{acc}$ while higher mass galaxies do not.
		Overmassive BHs in galaxies below $M\textsubscript{star} < 10^{10} M_\odot$ tend to fall above the observed relations.
		Notably, BHs that are considered overmassive or undermassive in $M\textsubscript{BH}$ are typically overmassive or undermassive in $M\textsubscript{acc}$ as well.
        Although it is unclear how much the BH seed mass affects growth, \Rom{} is capable of producing many BHs with growth consistent with current observational constraints by $z=0.05$.

	\subsection{BH Growth Modes} \label{subsec:results2}

		\begin{figure}
			\plotone{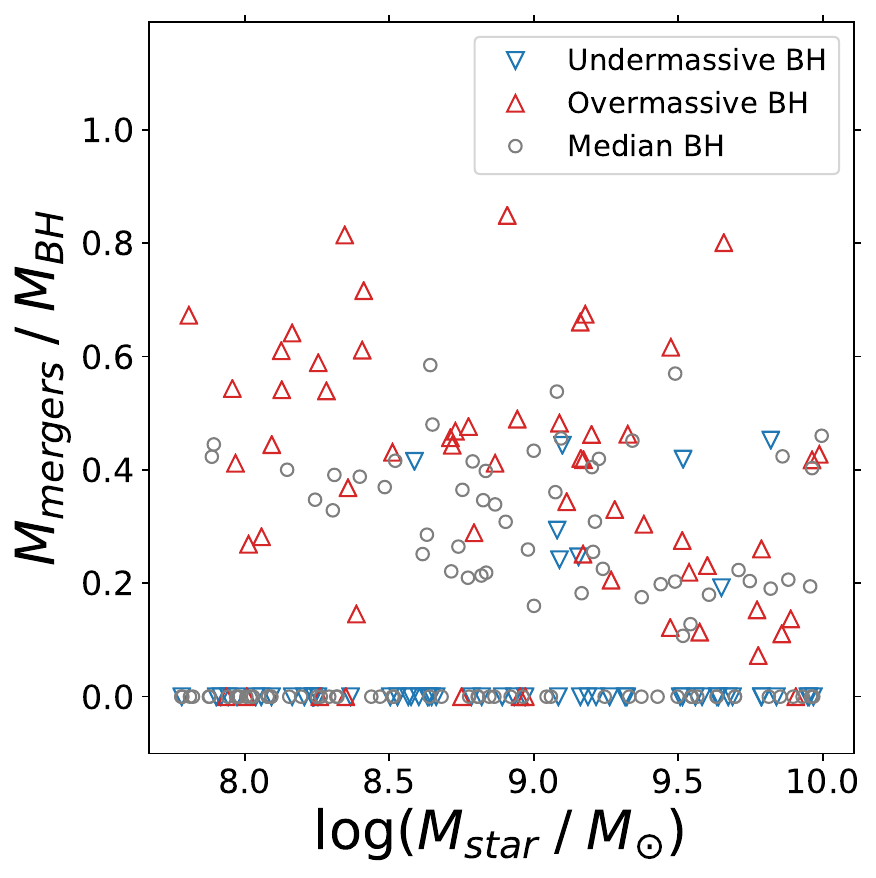}
			\caption{Fractional growth of BHs through mergers versus stellar mass, at $z=0.05$.
				Galaxies are colored by whether the BH is overmassive (red triangles), undermassive (blue inverted triangles), or median (grey circles).
				Nearly all undermassive BHs grow solely through accretion onto the main progenitor, never having experienced a BH merger.
				Overmassive and many median BHs grow through a combination of mergers and accretion onto the main progenitor.
				BHs in hosts with stellar masses $M\textsubscript{star} \lesssim 10^{9} M_\odot$ tend to have their growth dominated by mergers, while those in higher mass hosts tend to be dominated by accretion onto the main progenitor.
				\label{fractional-growth}}
		\end{figure}

		\begin{figure*}
			\plotone{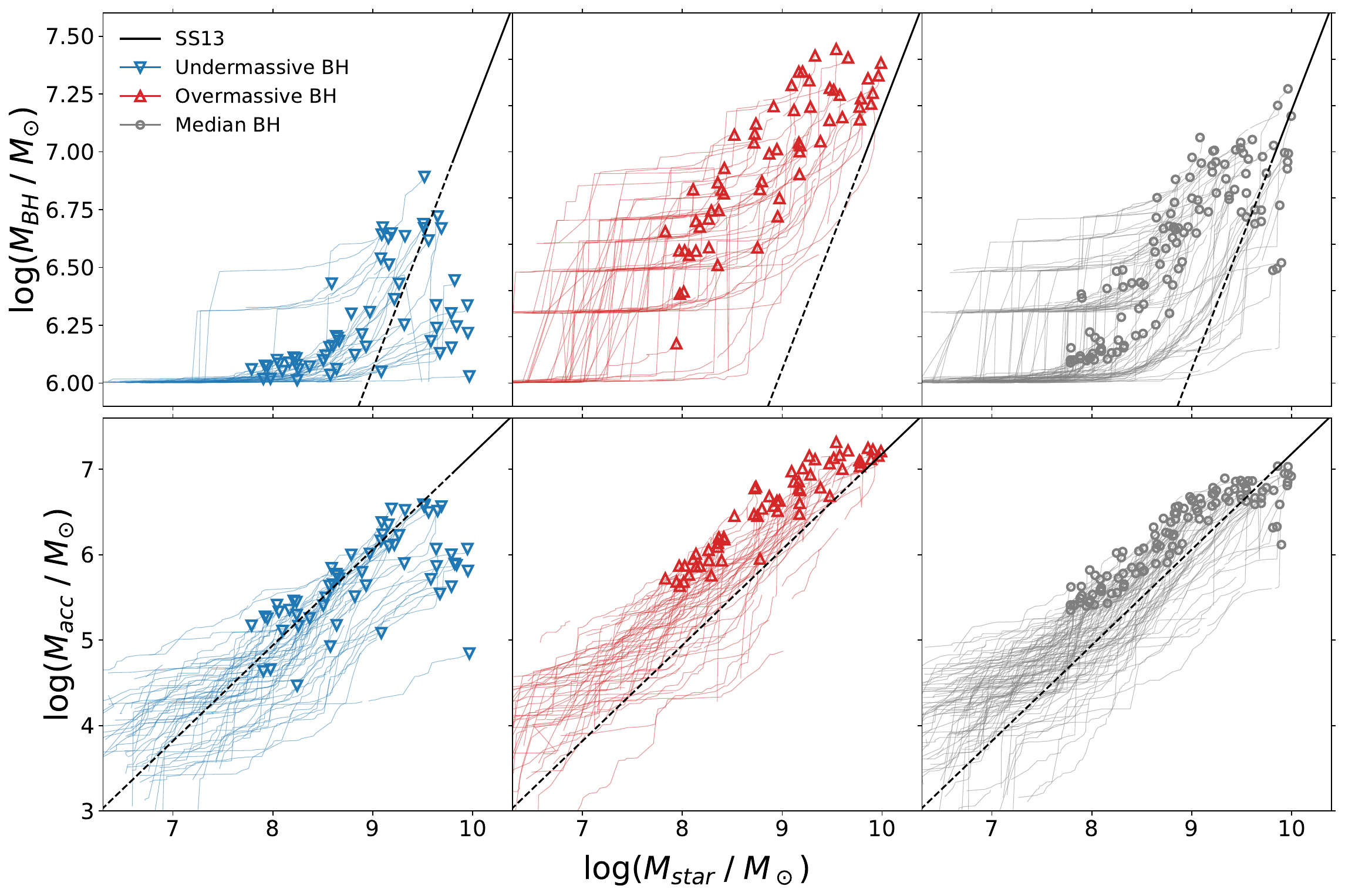}
			\caption{Evolution of BHs and their host galaxies, separated by undermassive (left), overmassive (center), and median BHs (right).
			The black line indicates the~\citet{Schramm2013} relation, where the dashed portion indicates the extrapolated relation.
			We plot both the growth of total BH mass and total BH accretion.
			\textit{Top}: Evolution onto the $z=0.05$ $M\textsubscript{BH} - M\textsubscript{star}$ relation.
			\textit{Bottom}: Evolution onto the $z=0.05$ $M\textsubscript{acc} - M\textsubscript{star}$ relation.
			Hosts of undermassive BHs tend to build up their stars more rapidly than their BHs, evolving onto the $z=0.05$ relation by growing stars first then BHs later.
			Hosts of overmassive BHs tend to build up their BHs rapidly before growing in stellar mass onto the $z=0.05$ relation.
			Median BHs tend to grow closer along the extrapolated SS13 relation.
			\label{Mbh-Mstar-evolution}}
		\end{figure*}

		Understanding the source of scatter in the $M\textsubscript{BH} - M\textsubscript{star}$ relation requires first understanding how the BHs evolved to the present day.
        BHs grow through BH-BH mergers as well as through accretion of gas particles.
		We trace the growth history of each BH and calculate the growth through mergers onto the main progenitor, $M\textsubscript{mergers}$.

        Figure~\ref{fractional-growth} shows the fraction of total BH mass grown via mergers by $z=0.05$, versus the host stellar mass.
        By $z=0.05$, only $8\%$ of undermassive BHs have ever undergone a BH merger, and instead the vast majority grow solely through accretion.
        On the other hand, $95\%$ of overmassive BHs have undergone at least one BH merger, and grow through a mixture of BH mergers and accretion at higher rates than undermassive BHs.
        The primary mode of growth for overmassive BHs depends on host stellar mass, where those found in hosts below $M\textsubscript{star} \lesssim 10^{9} M_\odot$ grow primarily through BH mergers, and those found in hosts above $M\textsubscript{star} \gtrsim 10^{9} M_\odot$ grow primarily through accretion onto the main BH progenitor.

		We next turn to how BHs evolve relative to the stellar mass of their hosts.
		Figure~\ref{Mbh-Mstar-evolution} shows the evolution of undermassive, overmassive, and median BHs and their hosts onto both the $z=0.05$ $M\textsubscript{BH} - M\textsubscript{star}$ relation and $M\textsubscript{acc} - M\textsubscript{star}$ relation.
		The black line indicates the SS13 locally observed relation, where the dashed portion indicates a linear extrapolation.
        Regardless of how we frame BH growth, we find fundamental differences in growth histories between undermassive and overmassive BH hosts.
        Undermassive BHs tend to evolve onto the $z=0.05$ relation by building up $M\textsubscript{BH}$ only after the host has built up its stars.
	    On the other hand, overmassive BHs hosts tend to either grow their BHs before their stellar mass, or grow both in tandem.
		The differences in $M\textsubscript{BH} - M\textsubscript{star}$ evolution strongly suggest overmassive and undermassive BH hosts may build up their stars and dark matter in fundamentally different ways.

	\subsection{Galaxy and Halo Growth} \label{subsec:results3}

		With the growth histories of the BHs in hand, we now study the environments in which they formed and reside.
		We trace the structural evolution of stars and dark matter in the BH hosts across cosmic time.
		In the following analysis we consider the \numNoBH{} isolated dwarf galaxies in \Rom{} which do not host any BHs alongside the \numBH{} which do host BHs, in order to better understand the role of BHs in structural evolution.

		\citet{Munshi2013} identify a systematic overestimate in dark matter only (DMO) simulation halo masses when compared to simulations that include baryon physics and outflows.
		When comparing with results from DMO simulations, we adjust halo masses such that $M\textsubscript{200, sim} = 0.8 M\textsubscript{200,DMO}$ for halos between masses $M\textsubscript{200} = 10^8 - 10^{12} M_\odot$.

		\subsubsection{Stellar Mass - Halo Mass Relation} \label{Results31}

			\begin{figure*}
				\epsscale{1.0}
				\plotone{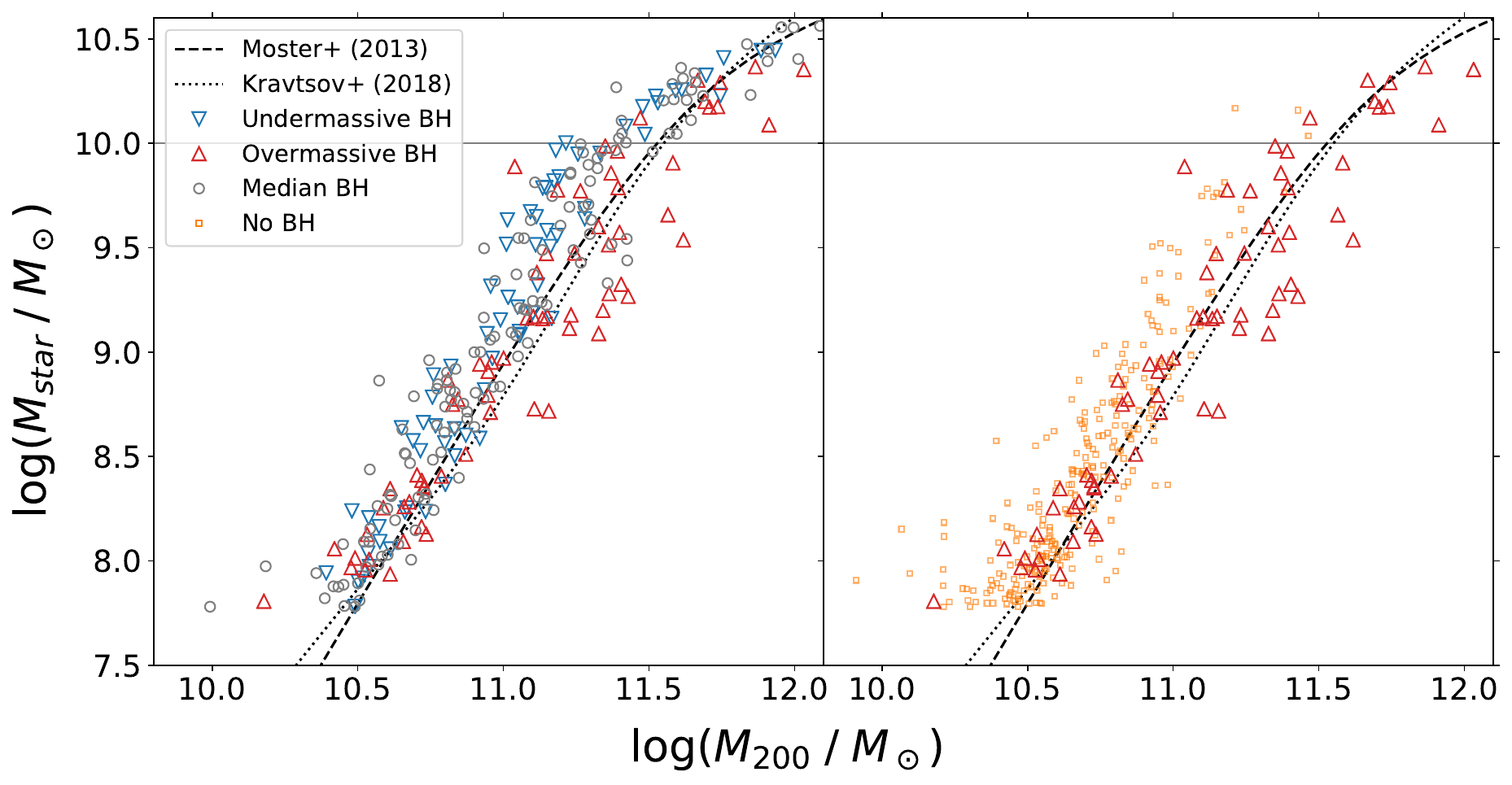}
				\caption{The $z=0.05$ stellar mass versus halo mass relation for isolated \Rom{} galaxies.
					Stellar and halo masses are adjusted using corrections from \citet{Munshi2013}.
				    We compare with abundance matching estimates from~\citet{Moster2013} (black dashed) and~\citet{Kravtsov2018} (black dotted).
					We limit the axes to focus on dwarf galaxy masses, and mark the $M\textsubscript{star} = 10^{10} M_{\odot}$ dwarf galaxy boundary (grey solid).
					\textit{Left}: Comparison of overmassive (red triangles), undermassive (blue inverted triangles), and median (grey circles) BH hosts.
					Above $M\textsubscript{200} > 10^{11} M_\odot$, overmassive BH hosts tend to sit at lower stellar masses than expected for the corresponding halo mass.
					Undermassive and median BHs tend to sit along or slightly above abundance matching estimates of the SMHM relation.
					\textit{Right}: Comparison of overmassive BH hosts with isolated galaxies that do not host a BH (orange squares).
					Galaxies without BHs exhibit a similar relation to galaxies hosting undermassive BHs, sitting along or slightly above both abundance matching estimates and overmassive BHs hosts.
					\label{SMHM}}
			\end{figure*}

			\begin{figure*}
				\epsscale{1.0}
				\plottwo{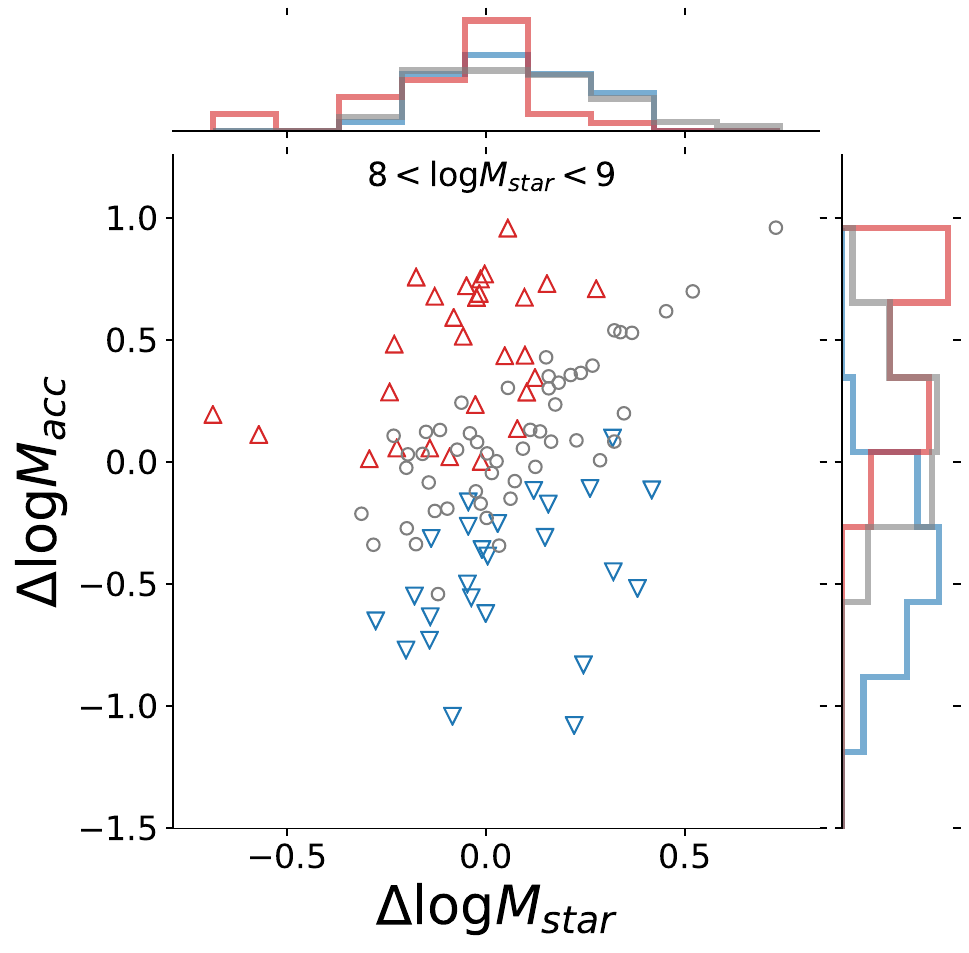}{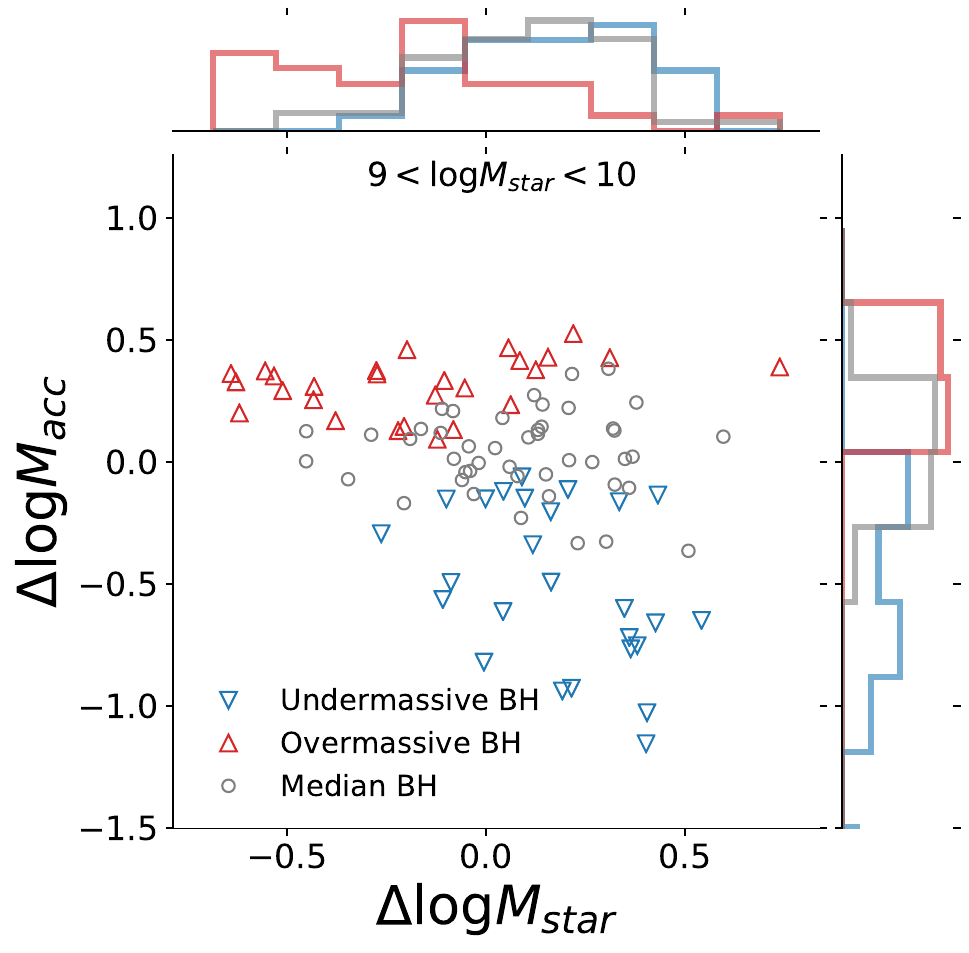}
				\caption{The residual quantities $\Delta \log M\textsubscript{acc}$ versus $\Delta \log M\textsubscript{star}$ (see text for definitions), split by galaxies with stellar mass $10^{8} < M\textsubscript{star} < 10^{9} M_\odot$ (left) and $10^{9} < M\textsubscript{star} < 10^{10} M_\odot$(right).
					Points are colored by whether the hosted BH is overmassive (red triangles), undermassive (blue inverted triangles), or median (grey circles).
					We also include normalized marginal histograms for each classification.
					\textit{Left}: Hosts of undermassive and overmassive BHs show identical distributions of $\Delta \log M\textsubscript{star}$.
					\textit{Right}: Galaxies with higher $\Delta \log M\textsubscript{acc}$ tend to be found at lower $\Delta \log M\textsubscript{star}$, and vice versa.
					Overmassive BH hosts tend to have the lowest $\Delta \log M\textsubscript{star}$.
					\label{Mbh-Mstar-residuals}}
			\end{figure*}

			Figure~\ref{SMHM} shows the $z=0.05$ stellar mass - halo mass (SMHM) relation for all isolated \Rom{} galaxies, with masses adjusted with corrections from~\citet{Munshi2013}.
			Points are colored by whether the hosted BH is overmassive, undermassive, median, or if the galaxy does not host a BH\@.
			A similar figure of the SMHM relation in \Rom{} for all central halos can be found in~\citet{Tremmel2017}.
			We mark the $M\textsubscript{star} = 10^{10} M_\odot$ dwarf galaxy boundary, and limit the axes to focus on low-mass galaxies.
			We include abundance matching estimates from~\citet{Moster2013} and~\citet{Kravtsov2018} for reference.
			Above $M\textsubscript{200} > 10^{11} M_\odot$, overmassive BHs tend to be found in halos with lower stellar masses than expected for their halo mass.
			Undermassive BHs instead tend to be found in halos with higher stellar masses than expected for their halo mass.
			Undermassive and median BH hosts in particular tend to sit along or above abundance matching estimates of the SMHM relation.
			Galaxies without BHs follow a similar relation to undermassive BH hosts, and above $M\textsubscript{200} > 10^{11} M_\odot$ sit at higher $M\textsubscript{star}$ than overmassive BH hosts at a given halo mass.

			To better quantify the connection between scatter in $M\textsubscript{acc}$ and scatter in $M\textsubscript{star}$, we define the residual quantities $\Delta \log M\textsubscript{acc}$ and $\Delta \log M\textsubscript{star}$.
			We fit a smoothing spline to the median SMHM relation for all isolated \Rom{} galaxies, then calculate $\Delta \log M\textsubscript{star}$ as the residual from the median $\log M\textsubscript{star}$ for a given halo mass.
			We similarly fit a spline to the median $M\textsubscript{BH} - M\textsubscript{star}$ relation and use the previous fit to find the expected $M\textsubscript{acc}$ for a given halo mass.
			We then calculate $\Delta \log M\textsubscript{acc}$ as the residual from the median.
			Figure~\ref{Mbh-Mstar-residuals} shows $\Delta \log M\textsubscript{acc}$ versus $\Delta \log M\textsubscript{star}$, split between galaxies with stellar mass $10^{8} < M\textsubscript{star} < 10^{9} M_\odot$ and $10^{9} < M\textsubscript{star} < 10^{10} M_\odot$.
			We split our sample in this way to better isolate resolution effects at the lowest masses.
			We distinguish between hosts of overmassive, undermassive, and median BHs.
			Positive / negative $\Delta \log M\textsubscript{acc}$ roughly correspond with overmassive / undermassive BHs, respectively.

			We find that overmassive BHs tend to be found in halos with fewer stars than expected from the median, while undermassive BHs are found in halos with more stars than expected.
			For galaxies with $10^{9} < M\textsubscript{star} < 10^{10} M_\odot$, overmassive BHs tend to be found at lower $\Delta \log M\textsubscript{star}$ and higher $\Delta \log M\textsubscript{acc}$ than their undermassive counterparts.
			Galaxies with $10^{8} < M\textsubscript{star} < 10^{9} M_\odot$ show little difference in $\Delta \log M\textsubscript{star}$ between overmassive and undermassive BHs.
			This result implies that BH accretion, and hence feedback, may suppress star formation in isolated galaxies between $10^{9} < M\textsubscript{star} < 10^{10} M_\odot$.

		\subsubsection{Structural Evolution} \label{Results32}

			\begin{figure*}
				\epsscale{0.8}
				\plotone{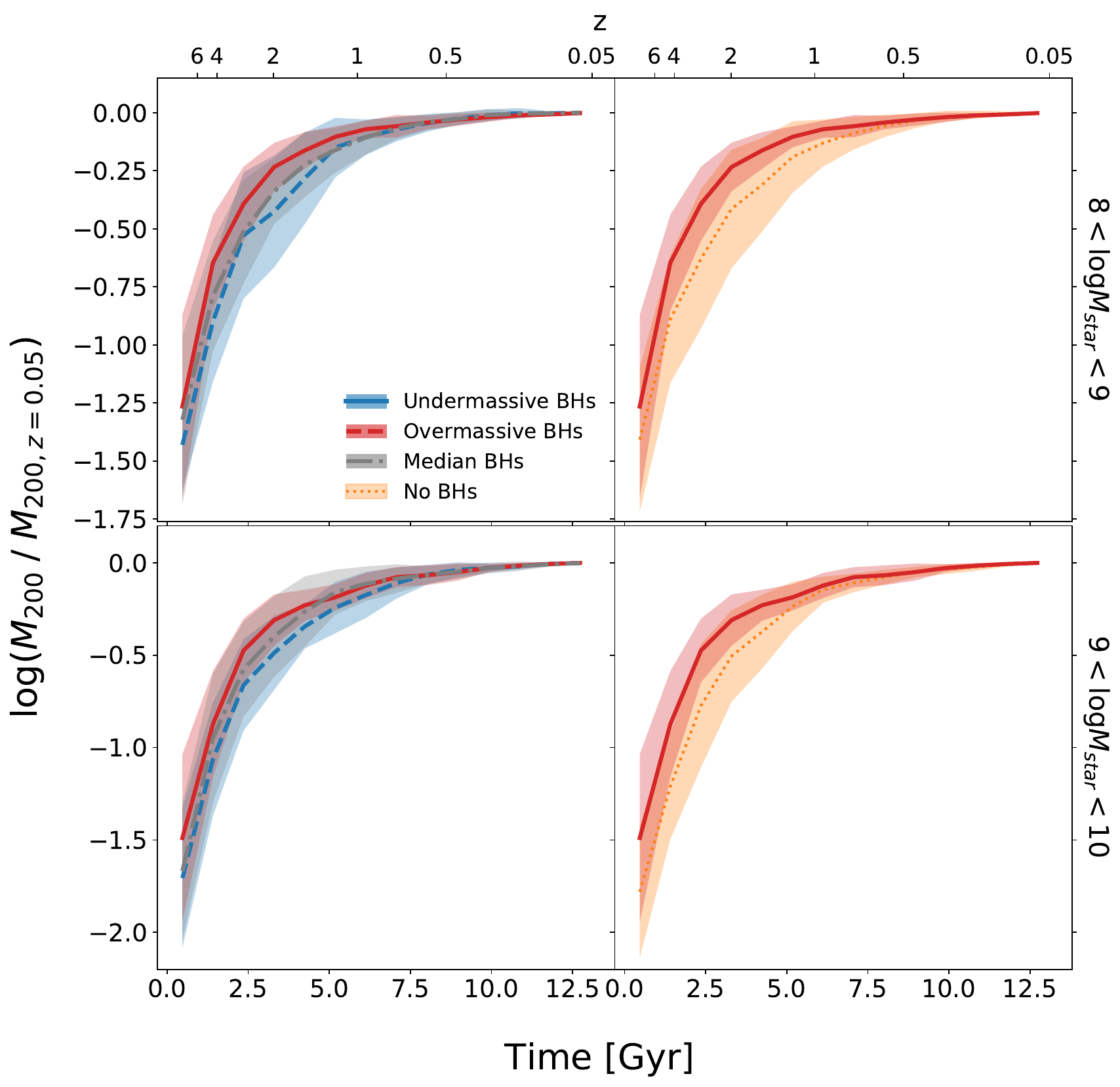}
				\caption{The median time evolution of halo mass of the main halo progenitor, scaled by the final $z=0.05$ halo mass.
					The left panels compare the halo mass evolution of hosts of undermassive (blue dashed), overmassive (red solid), and median (grey dash dot) BHs.
					The right panels compare the halo mass evolution of isolated galaxies without BHs (orange dotted) to the evolution of overmassive BH hosts.
					The top panels show galaxies with stellar mass $8 < M\textsubscript{star} < 9$, while the bottom panels show galaxies with stellar mass $9 < M\textsubscript{star} < 10$.
					Shaded regions indicate $1\sigma$ scatter in evolutionary tracks.
					Regardless of final stellar mass, hosts of overmassive BHs tend to build up their halo mass before hosts of undermassive BHs, as well as before galaxies without BHs.
					Undermassive BHs show a similar delay as galaxies without BHs in growing their halos.
					\label{M200-evolution}
				}
			\end{figure*}

			\begin{figure*}
				\epsscale{0.8}
				\plotone{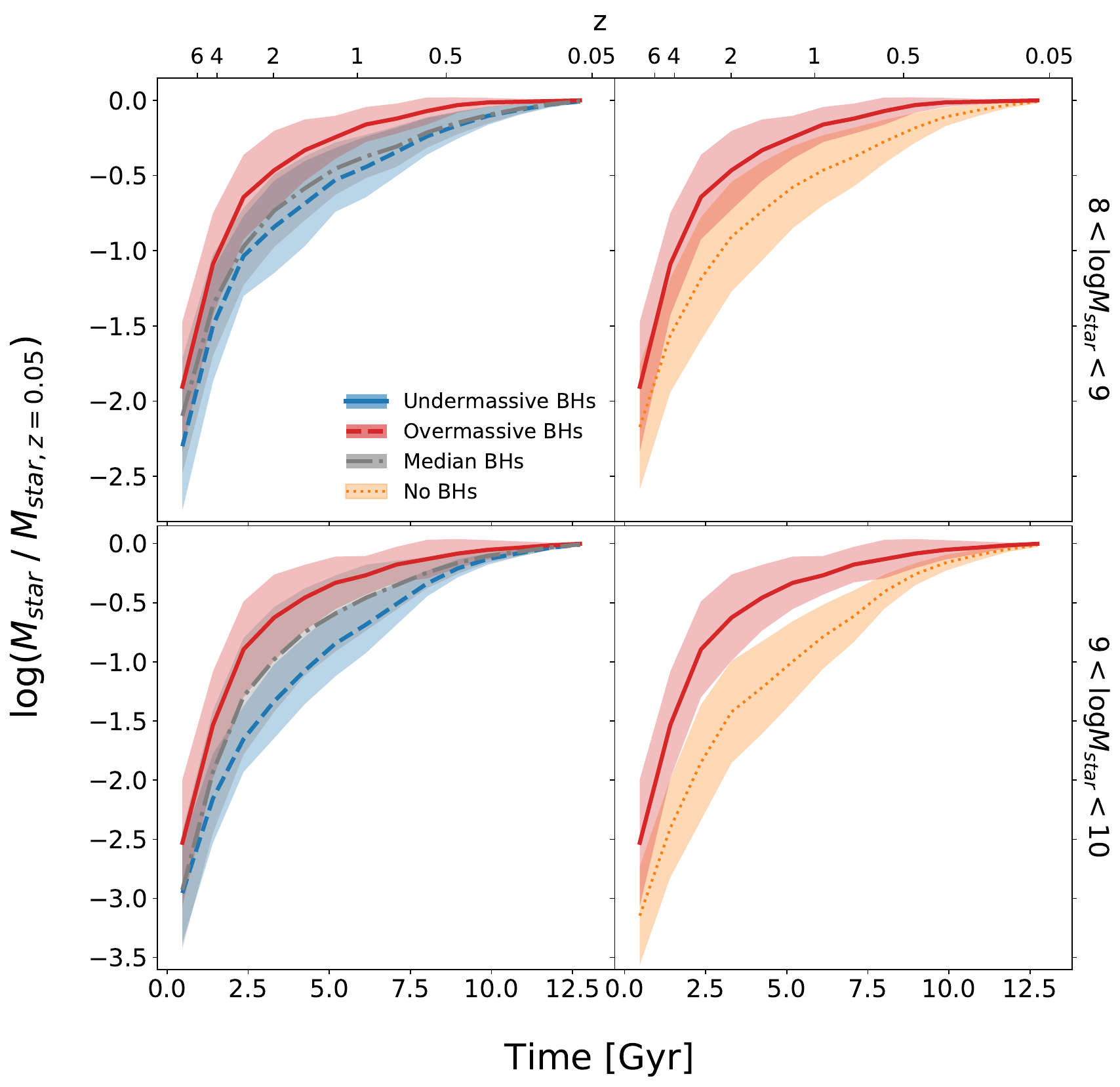}
				\caption{The median time evolution of stellar mass, scaled by the final $z=0.05$ stellar mass.
					The left panels compare the stellar mass evolution of hosts of undermassive (blue dashed), overmassive (red solid), and median (grey dash dot) BHs.
					The right panels compare the stellar mass evolution of isolated galaxies without BHs (orange dotted) to the evolution of overmassive BH hosts.
					The top panels show galaxies with stellar mass $10^{8} < M\textsubscript{star} < 10^{9} M_\odot$, while the bottom panels show galaxies with stellar mass $10^{9} < M\textsubscript{star} < 10^{10} M_\odot$.
					Shaded regions indicate $1\sigma$ scatter in evolutionary tracks.
					Hosts of overmassive BHs build up their stellar mass a few Gyr before hosts of undermassive BHs, as well as before galaxies without BHs.
					The differences in growth histories are most apparent in galaxies with final stellar masses $10^{9} < M\textsubscript{star} < 10^{10} M_\odot$.
					Undermassive BHs show a similar delay as galaxies without BHs in growing their stellar mass.
					\label{Mstar-evolution}
				}
			\end{figure*}

			\begin{figure*}
				\epsscale{0.8}
				\plotone{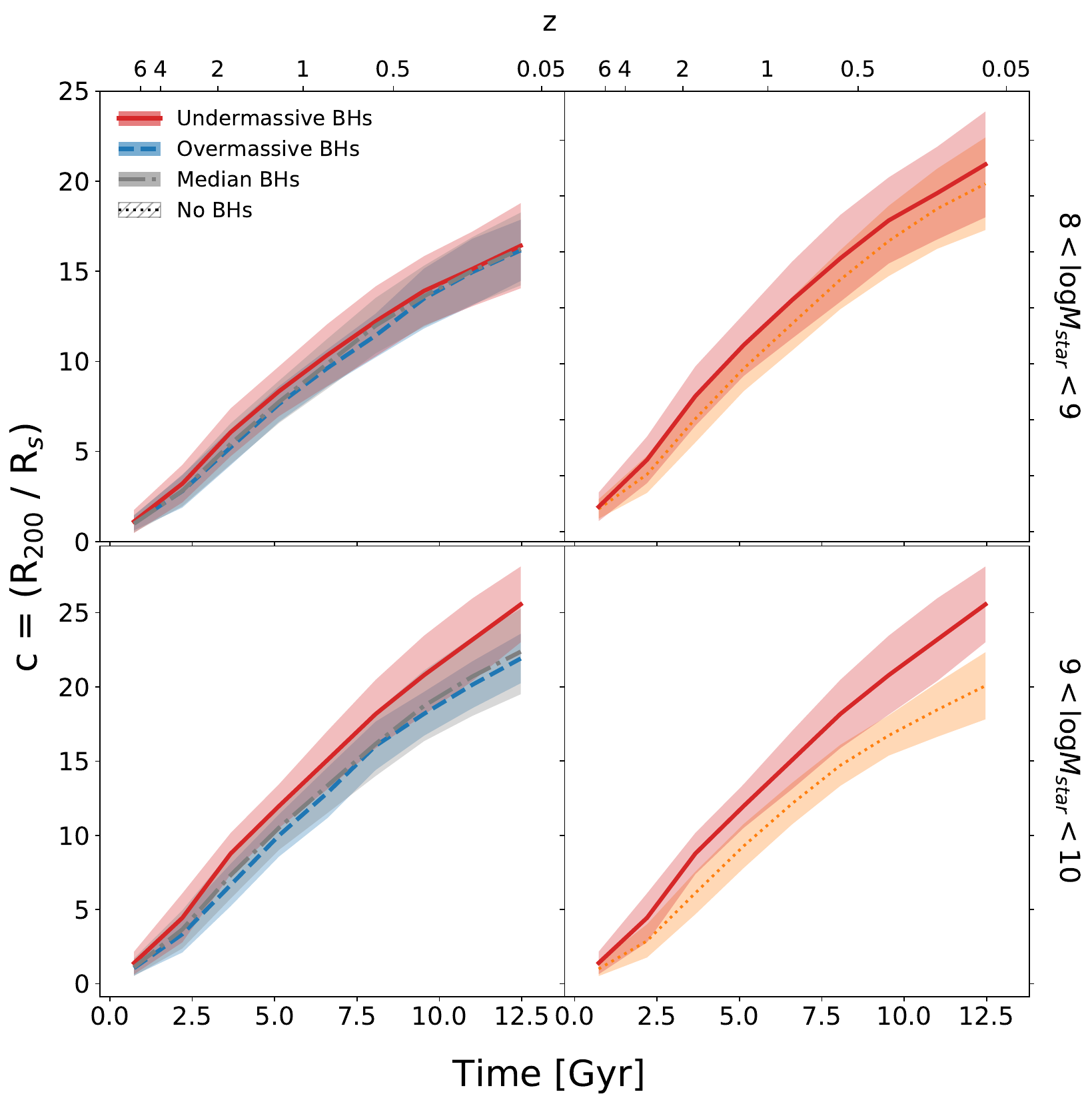}
				\caption{The median time evolution of halo concentration 	of the main halo progenitor. Ordering of panels and 	legend as in Figure \ref{M200-evolution}.
					For halos with final stellar mass $9 < \log M\textsubscript{star} < 10$, those that form overmassive BHs are more concentrated than their undermassive counterparts, and are more concentrated than halos without BHs by $z=0.05$.
					Halos with final stellar mass $8 < \log M\textsubscript{star} < 9$ show little variation in halo concentrations across time.
					\label{conc-evolution}
				}
			\end{figure*}

			\begin{figure*}
				\epsscale{0.8}
				\plotone{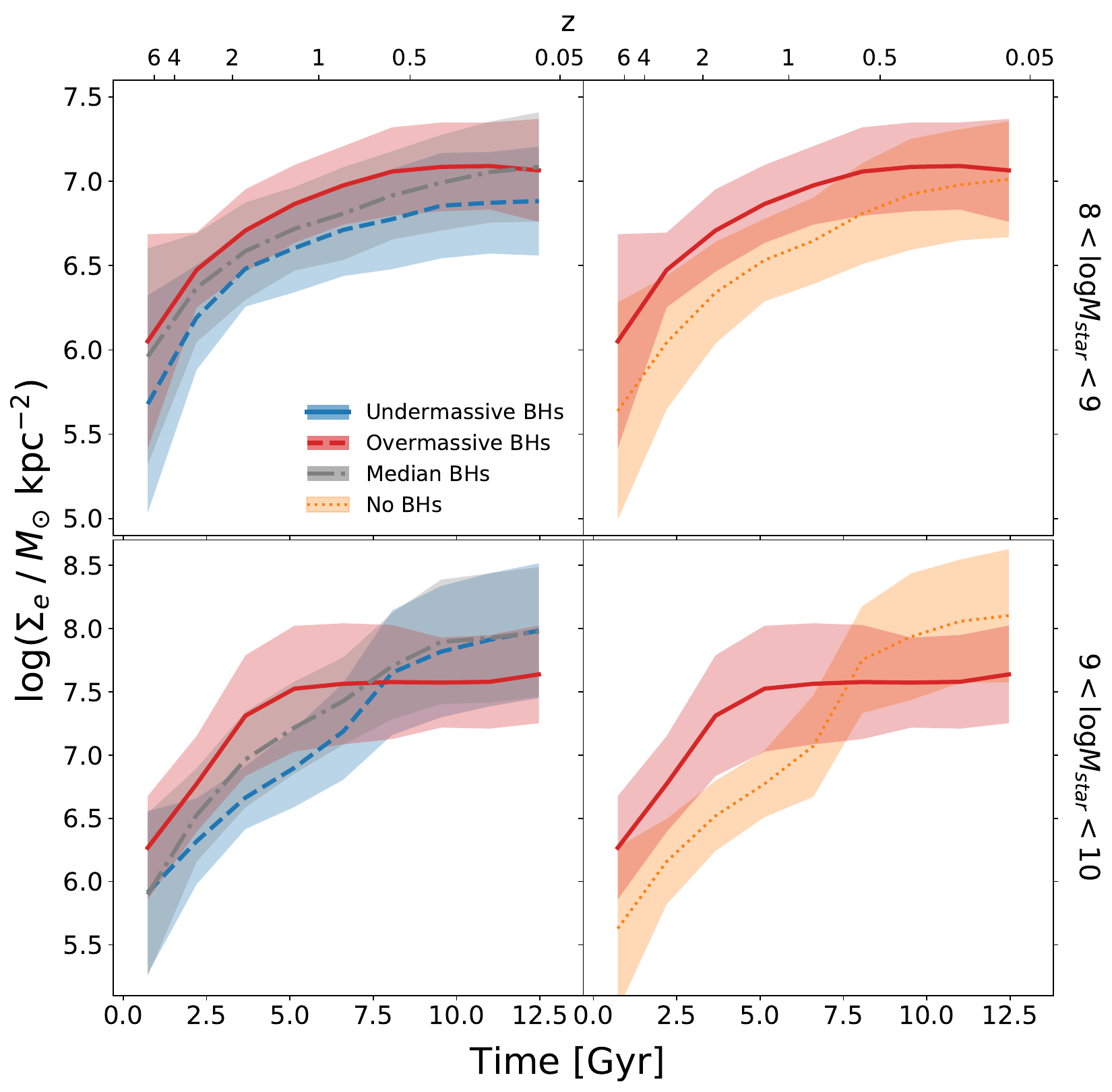}
				\caption{The median time evolution of central stellar mass surface density (see text for definition of $\Sigma\textsubscript{e}$).
					The left panels compare the evolution of hosts of undermassive (blue dashed), overmassive (red solid), and median (grey dash dot) BHs.
					The right panels compare the evolution of isolated galaxies without BHs (orange dotted) to the evolution of overmassive BH hosts.
					The top panels show galaxies with stellar mass $10^{8} < M\textsubscript{star} < 10^{9} M_\odot$, while the bottom panels show galaxies with stellar mass $10^{9} < M\textsubscript{star} < 10^{10} M_\odot$.
					Shaded regions indicate $1\sigma$ scatter in evolutionary tracks.
					Overmassive BH hosts grow stellar mass in their central regions more quickly than hosts of undermassive BHs, as well as more quickly than galaxies without BHs.
					Between $10^{8} < M\textsubscript{star} < 10^{9} M_\odot$, galaxies reach the same central densities by $z=0.05$.
					Between $10^{9} < M\textsubscript{star} < 10^{10} M_\odot$, overmassive BH hosts stop growing in $\Sigma\textsubscript{e}$ around $z\sim1$.
					Undermassive BH hosts and galaxies without BHs in this mass range continue growing, reaching higher $\Sigma\textsubscript{e}$ at late times than overmassive BH hosts.
					\label{Sigma-evolution}
				}
			\end{figure*}
			We now turn to the impact of BHs on the structural evolution of dark matter and stars.
			We find that the undermassive / overmassive nature of a BH is directly tied to the structural evolution of its host.
			Figure~\ref{M200-evolution} shows the evolution of $M\textsubscript{200}$ of the main halo progenitor, scaled by $M\textsubscript{200}$ at $z=0.05$, distinguishing between undermassive, overmassive, and median BH hosts.
			We also include a comparison between galaxies hosting overmassive BHs and galaxies without BHs.
			We distinguish between galaxies with $z=0.05$ stellar mass $10^{8} < M\textsubscript{star} < 10^{9} M_\odot$ and $10^{9} < M\textsubscript{star} < 10^{10} M_\odot$.
			Overmassive BH hosts tend to build up their halos earlier than both hosts of undermassive BHs and galaxies without BHs.
			The delay in halo growth is most pronounced in dwarf galaxies above $M\textsubscript{star} > 10^{9} M_\odot$ at $z=0.05$.

			Figure~\ref{Mstar-evolution} similarly shows the median cumulative star formation history for each class of BH hosts.
			Overmassive BH hosts build up their stellar mass much more rapidly than both hosts of undermassive BHs and galaxies without BHs.
			As with halo mass, the differences in stellar mass evolution are most pronounced in dwarf galaxies above $M\textsubscript{star} > 10^{9} M_\odot$ at $z=0.05$.

			Further analysis shows that the formation of overmassive / undermassive BHs can likely be attributed to differences in halo assembly times.
			We fit each halo density profile with a flexible 5-parameter model \citep{Dekel2017}:
			\begin{align}
				\rho(r) = \frac{\rho_c}{x^\alpha(1 + x^{1/\beta})^{\beta(\gamma - \alpha)}},
			\end{align}
			where $\alpha$ and $\gamma$ are respectively the inner and outer profile slopes, $\beta$ is the intermediate shape parameter, and the radius is scaled by the scale radius, $R\textsubscript{s}$, which is related to the halo concentration, $c$:
			\begin{align}
				x = \frac{r}{R_s},\quad
				c = \frac{R\textsubscript{200}}{R\textsubscript{s}}.
			\end{align}

			Figure \ref{conc-evolution} shows the evolution of halo concentration over time.
			Overmassive BH hosts have higher halo concentrations than their undermassive counterparts see \cite[see][and references therein]{Maccio2008,Zhao2009}, as well as halos without BHs.
			A two-sample, two-sided Kolmogorov-Smirnov (KS) test on overmassive versus undermassive BH host concentrations yields a KS statistic $D_{n,m} = 0.29$ and allows us to reject the null hypothesis at the $0.02$ level that the halos are drawn from the same distribution of concentrations. Similarly, a KS test on overmassive BH hosts versus halos without BHs yields a KS statistic $D_{n,m} = 0.45$ and allows us to reject the null hypothesis at the $8\times10^{-8}$ level.
			
			The build up of stars prior to BH growth, the lack of BH mergers in undermassive BH hosts, and the lower halo concentrations all suggest undermassive BHs initially formed in environments with a lower abundance of cold gas than is necessary to seed multiple BHs.
			In contrast, overmassive BHs were likely initially seeded in environments with an abundance of cold gas, where BHs had a higher likelihood to merge and accrete efficiently.

			There is strong evidence that the central regions of massive galaxies are most affected by the presence of a central BH~\citep{Cheung2012,Barro2017,Choi2018}.
			We trace the evolution of stars within the central regions of the host galaxy and search for a connection to the central BH\@.

			We define the stellar mass surface density within the stellar half-light radius, $r_e$:
				\begin{align}
					\Sigma\textsubscript{e} = \frac{M\textsubscript{star}(< r_e)}{\pi r_e^2},
				\end{align}
			where we calculate $r\textsubscript{e}$ by fitting a S\'ersic profile to the projected face-on $V$-band surface brightness profile, with a surface brightness cutoff of $32$ mag arcsec$^{-2}$~\citep{Abraham2014}.

			\citet{Cheung2012} find that stellar density within the central $1$ kpc, $\Sigma_1$,
			robustly follows quenching in local galaxies above $M\textsubscript{star} \gtrsim 10^{8} M_\odot$.
			\citet{Chen2019} build a schematic model that finds galaxies begin to rapidly quench once they evolve over a boundary in $\Sigma\textsubscript{1} - M\textsubscript{star}$ space.
			Many \Rom{} dwarf galaxies fall below $r\textsubscript{e} < 1$ kpc at early times, hence we use $r_e$ to consistently define a central region across time.
			Both~\citet{Franx2008} and~\citet{Barro2013} find a strong relationship between $\Sigma\textsubscript{e}$ and both the star formation rate and total stellar mass out to $z\sim 3.5$.
			Hence, tracing the evolution of $\Sigma\textsubscript{e}$ while distinguishing between hosts of overmassive and undermassive BHs can give insight into the effects of BH growth on central star formation.

			Figure~\ref{Sigma-evolution} shows the evolution of $\Sigma\textsubscript{e}$ across time for galaxies with undermassive, overmassive, and median BHs, as well as for galaxies without BHs.
			We distinguish between galaxies with $z=0.05$ stellar mass $10^{8} < M\textsubscript{star} < 10^{9} M_\odot$ and $10^{9} < M\textsubscript{star} < 10^{10} M_\odot$.
			Similar to the buildup of total stellar mass, overmassive BH hosts buildup their central stellar density more rapidly than both undermassive BH hosts and galaxies without BHs.
			However, above $M\textsubscript{star} > 10^{9} M_\odot$, overmassive BH hosts show suppression of $\Sigma\textsubscript{e}$ growth starting at redshift $z\sim2$.
			By $z\sim0.5$, many hosts of undermassive BHs and galaxies without BHs reach $0.5$ dex higher $\Sigma\textsubscript{e}$ than hosts of overmassive BHs.
			Hosts of undermassive BHs and galaxies without BHs may instead flatten in $\Sigma\textsubscript{e}$ at late times, around $z\sim0.1$.
			We confirm that these results remain qualitatively the same were we to use surface densities calculated within the central $1$ kpc rather than $r_e$.

			In short, overmassive BHs were first seeded in early-forming halos, building up their dark matter and stars rapidly in tandem with growth of the BH.
			Despite having higher halo concentrations, overmassive BH hosts have similar or lower central stellar density than their undermassive counterparts by late times, indicating a measure of star formation suppression within the central regions.
			Undermassive BH hosts instead follow nearly identical evolutionary tracks to galaxies without BHs, growing dark matter and stars later and exhibiting high central stellar densities at late times.
			~\citet{Dickey2019} find a similar relationship in isolated galaxies with stellar mass $10^{9} < M\textsubscript{star} < 10^{9.5} M_\odot$, where strong signatures of AGN correlate with an older stellar population in the host galaxy.
			~\citet{Li2019} similarly find overmassive BH hosts form earlier and have lower present-day star formation rates.

			\citet{Choi2018} find similar evolution of $\Sigma\textsubscript{e}$ in zoom-in simulations of galaxies with $M\textsubscript{star} > 10^{10.9} M_\odot$ run with and without AGN feedback.
			They find galaxies with AGN feedback build up $\Sigma\textsubscript{e}$ until $z\sim2$, after which $\Sigma\textsubscript{e}$ turns over and begins to decrease.
			The stellar cores become diffuse primarily through AGN-driven stellar mass-loss and gas mass-loss ``puffing-up'' the central region.
			They find the turnover in $\Sigma\textsubscript{e}$ is concurrent with quenching of star formation.
			Galaxies run without AGN feedback indefinitely increase their central stellar densities and do not experience the same level of quenching.
			Both~\citet{Guo2013} and~\citet{Barro2017} similarly observe stellar cores diffuse over time in CANDELS GOODS-S galaxies with stellar masses $10^{9} < M\textsubscript{star} < 10^{11.5} M_\odot$.

			Stagnation in central stellar density can come about in a few ways.
			Stellar evolution can eject mass from stars and return it to gas in the ISM~\citep{VanDokkum2014}.
			Mass-loss through stellar outflows directly competes with new star formation promoted by the increased gas mass~\citep{Kennicutt1994}.
			As a result~\citet{Choi2018} find this effect to contribute little to central stellar density suppression.

			Compaction events may occur through strongly dissipational, gas-rich mergers driving stars and gas toward the galaxy center.
			Conversely, gas-poor mergers can reduce the core stellar density by driving rapid size growth with little growth in mass~\citep{Hopkins2008,Nipoti2009,Covington2011,Oser2012,Hilz2013,Porter2014}.
			Overmassive BH hosts in \Rom{} are often found to have up to 4 orders of magnitude lower gas fractions relative to galaxies without BHs (see Section \S\ref{subsubsec:coldgas}) and hence may be subject to stellar diffusion through gas-poor mergers, though we find no evidence of major mergers driving rapid changes in stellar or gas profiles of overmassive or median BH hosts.
			
			There are other physical processes that have been commonly found to reduce central stellar density, but are unresolved in \Rom{}.
			Binary black hole systems may be capable of clearing out galaxy cores~\citep{Milosavljevic2001,Kormendy2009}, but BH scouring occurs on scales below the spatial resolution limit of \Rom{}~\citep{Rantala2017,Rantala2018}.
			Mass loss and heating driven by outflows and SNe may reduce the gravitational potential and in turn allow for outward stellar migration~\citep{Fan2008,Choi2018}, but \Rom{} does not resolve these effects on the central potential.
			
            Finally, heating and mass-loss driven by outflows and SNe can in turn drive gas outward, restrict gas cooling, and hence suppress star formation~\citep{Somerville2015}. Below we show that the differences in structural evolution between \Rom{} galaxies with overmassive BHs and other galaxies are likely due this BH feedback.
            
	\subsection{Impact of BHs on Stars and Gas} \label{subsec:results4}

		\subsubsection{Energy Injection by BHs}

			\begin{figure}
				\plotone{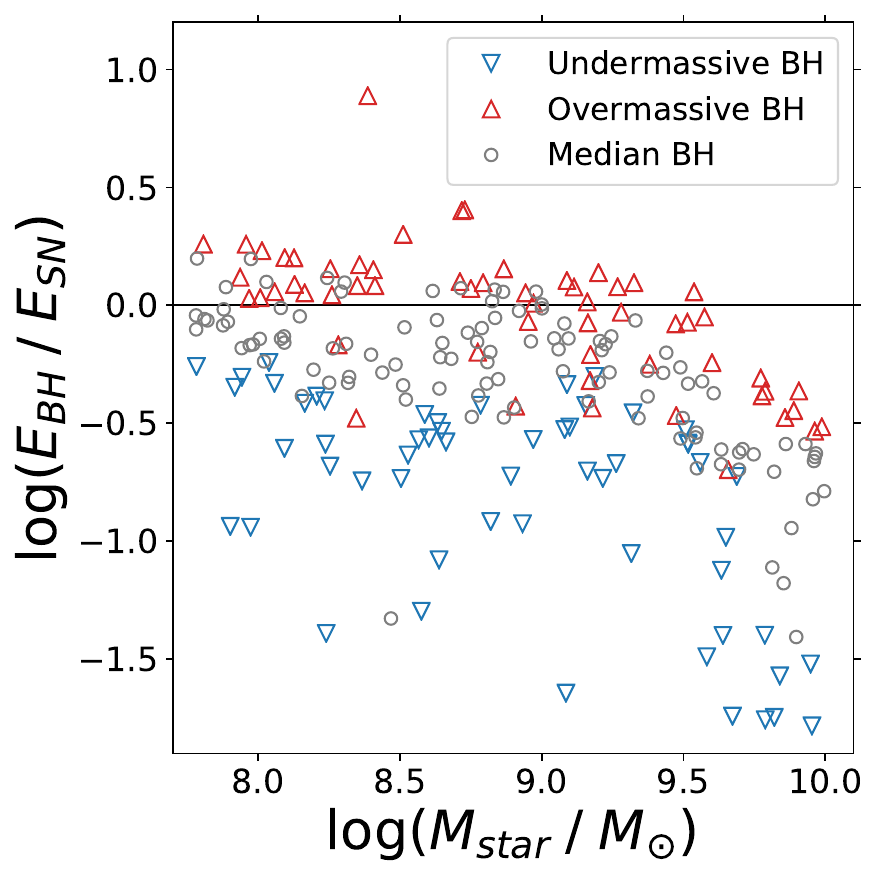}
				\caption{
				Ratio of the total energy injected by BHs to the total energy injected by SNe over cosmic time, versus the stellar mass of the host galaxy at $z=0.05$.
					Galaxies are colored by whether they host an overmassive (red triangles), undermassive (blue inverted triangles), or median (grey circles) BH.
					We mark the $E\textsubscript{BH} = E\textsubscript{SN}$ boundary (black line).
					Undermassive BH hosts are all dominated by energy injection from SNe.
					Regardless of host stellar mass, overmassive and median BHs are often capable of injecting more energy than SNe.
					\label{EBH-ESN}
				}
			\end{figure}

			We find that median and overmassive BHs are capable of injecting more energy than SNe into the surrounding gas of their hosts.
			We calculate total energy injection via BHs and SNe by first integrating their energy outputs across cosmic time.
			A set fraction of the energy output is injected into the surrounding interstellar medium (see Section~\ref{subsec:simulation-properties}).
			Figure~\ref{EBH-ESN} shows the ratio of energy injected by BHs to SNe, versus the host stellar mass.
			Many overmassive and median BHs are capable of injecting substantially more energy than SNe.
			On the other hand, undermassive BH hosts are all dominated by energy injection by SNe.
			At all stellar masses, overmassive BHs hosts are injected with $2-3$ more combined BH + SNe energy than undermassive BHs hosts.

			A higher $E\textsubscript{BH}/E\textsubscript{SN}$ suggests that outflows from BHs may more efficiently heat and drive gas than SNe outflows.
			It is important to note that, while BHs produce and inject copious amounts of energy, it is ultimately the feedback prescription that determines the effect on the host.
			Feedback models that inject kinetic energy typically drive winds at higher velocities than those that inject purely thermal energy~\citet{Choi2018}.

		\subsubsection{Cold Gas Depletion} \label{subsubsec:coldgas}

			\begin{figure}
				\plotone{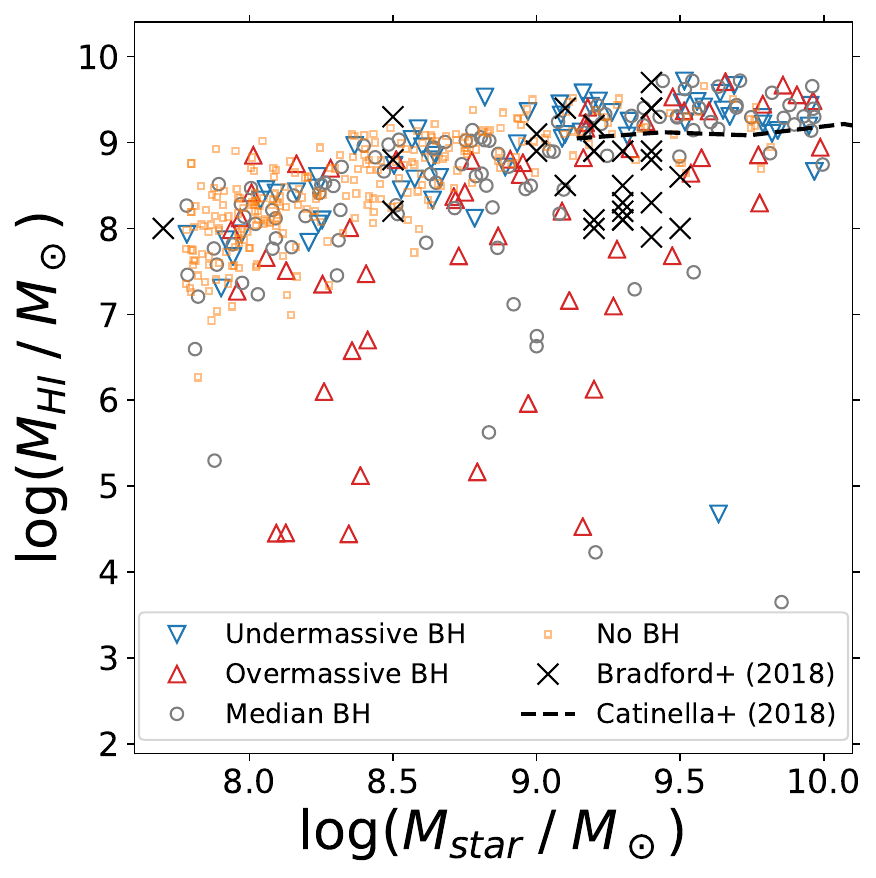}
				\caption{\HI gas mass versus stellar mass for hosts of overmassive (red triangles), undermassive (blue inverted triangles), and median (grey circles) BHs.
					Galaxies without BHs are also shown (orange squares).
					We compare with observed $M\textsubscript{HI} - M\textsubscript{star}$ relations from~\citet{Bradford2018} (black crosses) and~\citet{Catinella2018} (dashed black).
					Undermassive BH hosts and galaxies without BHs agree well with the observations.
					Overmassive BH hosts tend to have significantly less \HI gas than other galaxies at the same stellar mass.
					\label{MHI-Mstar}
				}
			\end{figure}

			Next we turn to the relationship between BHs and the amount of cold gas in the host galaxy.
			Figure~\ref{MHI-Mstar} shows the \HI gas mass versus stellar mass relation for isolated dwarf galaxies.
			We compare with observations from ~\citet{Bradford2018} and~\citet{Catinella2018}.
			Bradford et al.~combine a new set of $21$ cm observations with the \HI-selected ALFALFA survey~\cite{Haynes2011} to analyze gas depletion in local galaxies ($0.002 < z < 0.055$) with stellar mass $10^{7} < M\textsubscript{star} < 10^{9.5} M_\odot$.
			Catinella et al.~measure \HI content of local ($0.01 < z < 0.05$) stellar mass selected xGASS galaxies, with stellar masses $10^{9} < M\textsubscript{star} < 10^{11.5} M_\odot$.

			Hosts of undermassive BHs and galaxies without BHs follow the Catinella et al.~relation at the high mass end, and are consistent with non-depleted galaxies from Bradford et al.~across all stellar masses.
			Hosts of undermassive BHs and galaxies without BHs show little indication of significant cold gas depletion.
			On the other hand, many overmassive and median BH hosts exhibit lower $M\textsubscript{HI}$ than expected for their stellar mass, indicating a high degree of cold gas depletion.
			Although ALFALFA is most sensitive to \HI above $M\textsubscript{HI} \gtrsim 10^{7} M_\odot$, it is worth noting the extreme levels of cold gas depletion seen in overmassive BH hosts are often orders of magnitude below what is seen in either observational comparison sample.
			Bradford et al.~similarly find that isolated galaxies with stellar mass $10^{9.2} < M\textsubscript{star} < 10^{9.5} M_\odot$ with strong signatures of AGN tend to show a higher degree of gas-depletion than similar galaxies with weaker AGN signatures, though they find that this effect does not extend to more massive galaxies.
			Bradford et al.~do not rule out the ejection and heating of cold gas by unusually bursty and compact star formation activity.

		\subsubsection{Star Formation Quenching}

			\begin{figure}
				\plotone{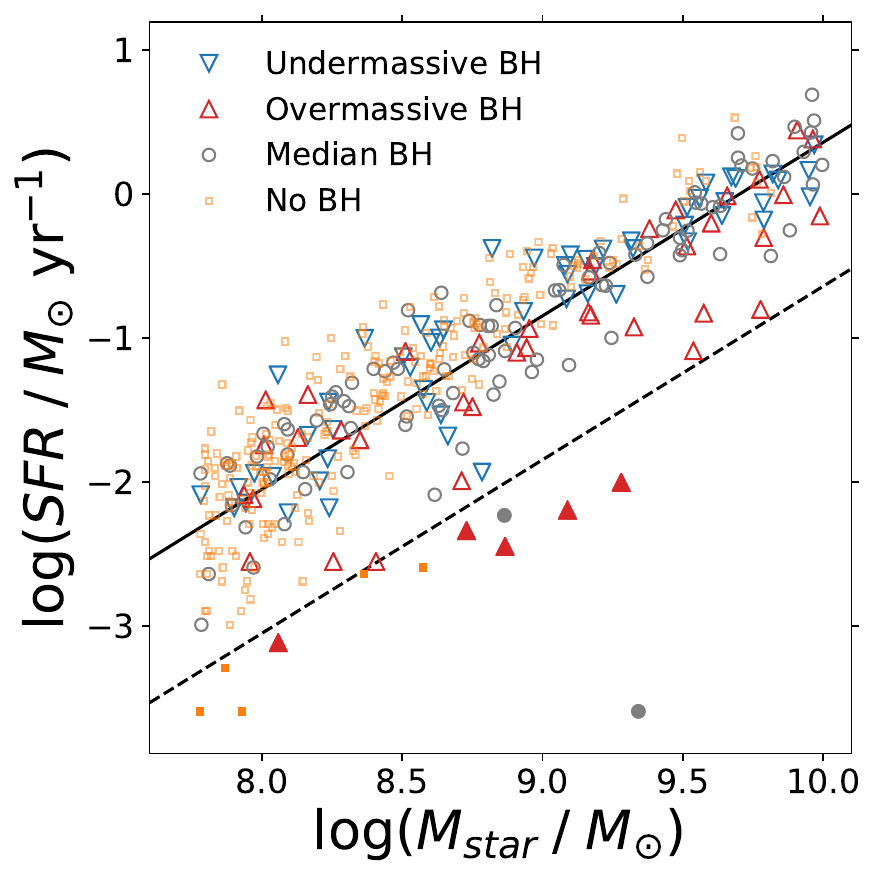}
				\caption{Star formation rate versus stellar mass for hosts of overmassive (red triangles), undermassive (blue inverted triangles), and median (grey circles) BHs.
					Galaxies without BHs are also shown (orange squares).
					We plot the $z=0$ \Rom{} main sequence as calculated by~\citet{Tremmel2019} (black solid), and the quenched boundary 1 dex below the main sequence (black dashed).
					Filled points indicate galaxies we consider quenched (see text for details).
					The majority of quenched galaxies with BHs host overmassive BHs.
					Galaxies without BHs can be quenched, but are all found at lower stellar masses, $M\textsubscript{star} < 10^{9} M_\odot$.
					\label{SFR-MS}
				}
			\end{figure}

			\citet{Tremmel2019} fit the \Rom{} star formation main sequence by calculating median values of the star formation rate (SFR) within $0.1$ dex bins of stellar mass between $10^{8} < M\textsubscript{star} < 10^{10}$, for galaxies considered relatively isolated by the same criteria we use in this work.
			They find a best fit of $\log(SFR) = 1.206 \times \log(M\textsubscript{star}) - 11.7$ at $z=0$.
			We define galaxies whose star formation rate falls a factor of $10$ below the main sequence to have quenched star formation.
			We calculate SFRs averaged over the previous $250$ Myr.

			Figure~\ref{SFR-MS} shows the $z=0.05$ star formation main sequence for \Rom{} dwarf galaxies.
			We distinguish between hosts of overmassive, undermassive, and median BHs, as well as isolated galaxies without BHs.
			Quenched galaxies are marked with filled points.
			We mark the~\citet{Tremmel2019} relation with a solid line.

			Quenching tends to occur in isolated dwarf galaxies that host overmassive or median BHs.
			We find $12$ quenched dwarf galaxies, $7$ of which host a BH\@.
			Above $M\textsubscript{star} \gtrsim 10^{8.6} M_\odot$, quenching only occurs in dwarf galaxies that host BHs.
			Quenched galaxies that host a BH tend to host overmassive BHs, regardless of host stellar mass.
			
			To help identify the source of quenching as internal or external, we track the interaction history of each galaxy of interest and estimate the tidal effects of close encounters.
			Following~\citet{Karachentsev1999}, we calculate tidal indices, $\Theta$, such that
			\begin{align}
				\Theta = \max\{\log(M_k/D_k^3)\} &- 11.75,\\
				&k = 1, 2, ..., N,
			\end{align}
			where $M$ and $D$ are the masses and 3D separations, respectively, of the $k$th closest halo.
			Negative values of $\Theta$ indicate no tidal disturbance of the main halo by nearby halos, while high positive values (we arbitrarily choose $\Theta > 5$) indicate significant tidal disturbance.

			Analyzing the encounter history of our quenched galaxies, there is $1$ undermassive BH host that is particularly close to the quenched boundary and further analysis reveals it was stripped of \HI gas and quenched immediately following tidal disturbance by a more massive halo.
			Similarly, the $2$ quenched median BH hosts both show evidence of such encounters followed by \HI depletion and quenching.
			Of the $5$ quenched overmassive BHs, $2$ exhibit high tidal indices from encounters with massive halos.
			Thus, quenching can occur in galaxies below $M\textsubscript{star} < 10^{8.6} M_\odot$ even if they do not host a BH\@, though all but $1$ show clear evidence of past encounters with a more massive galaxy and a subsequent drop in star formation.

			As discussed in~\citet{Tremmel2019}, our definition of quenched is different from some observations, such as~\citet{Wetzel2012} who adopt a flat threshold in specific SFR of $10^{-11}$ yr$^{-1}$.
			Regardless, our results change little when we instead use a flat specific SFR threshold of $10^{-11}$ yr$^{-1}$.

			Although \Rom{} uses a purely thermal feedback model, simulations have found success in using feedback models that incorporate both thermal feedback and mechanical feedback that efficiently drives high velocity winds.
			~\citet{Choi2015} find the inclusion of mechanical feedback more efficiently suppresses late-time star formation and produces AGN luminosities in line with observations.
			~\citet{Choi2017} find mixed thermal and mechanical AGN feedback models yield reasonable results for ex-situ and in-situ star formation, and realistic gas and stellar structural properties.
			~\citet{Weinberger2017} find that dual-mode AGN feedback for weakly/highly accreting BHs yields realistic star formation suppression, gas fractions, BH growth, and thermodynamic profiles.

			Our findings are in line with results from~\citet{Ricarte2019}, who find that isolated \Rom{} galaxies with $M\textsubscript{star} > 10^{9.5} M_\odot$ show signs of co-evolution with their central BH\@.
			Ricarte et al find that the BH accretion rate follows the SFR in star-forming galaxies, regardless of redshift, stellar mass, or large-scale environment.
			Further, they find such BHs grow in tandem with their host galaxies, eventually being confined to a line of constant $M\textsubscript{BH} / M\textsubscript{star}$.
			They suggest self-regulation of BH growth through feedback is a possible driver of co-evolution seen in isolated \Rom{} galaxies.

	\subsection{Caveats} \label{subsec:caveats}

		Our analysis of the role of BHs in dwarf galaxy evolution has a few caveats.
		Although we find correlations between BH properties and properties of the host galaxy, the precise effect of BH activity on dwarf galaxy evolution is unclear.
		We have not directly traced the effects of BHs on the surrounding environment (e.g., tracing outflows, tracking heating, turning off BH physics and rerunning the simulation), and hence we cannot say for certain how BHs may drive changes in their hosts within \Rom{}.
		Further, observations from \citet{Mezcua2019} find that dwarf galaxies can host powerful jets whose mechanical feedback may strongly impact the host star formation history, an effect which is not accounted for in \Rom{}.
		We find no evidence of starbursts in compact galaxies~\citep{Diamond-Stanic2012} as the dominant mechanism driving trends in star formation suppression, though the spatial resolution of \Rom{} likely restricts especially compact galaxies from forming.
		It appears BH feedback plays a larger role than often thought within dwarf galaxies~\citep{Martin-Navarro2018}, and it is likely both stellar and BH feedback together drive suppression of central stellar density and overall star formation in \Rom{} dwarf galaxies.
		Future work will further analyze the role of AGN in the evolution of dwarf galaxies, in particular how BH activity relates to suppression of star formation.

		As seen in Equation~\ref{Accretion Rate}, accretion onto BHs is sensitive to the BH mass.
		This is particularly important for two reasons: 1) some BHs in \Rom{} may unphysically merge immediately after seeding, and 2) the seed mass is likely too high in the lowest mass galaxies.
		Some BHs effectively form at higher masses than the seed mass due to seeding of multiple, clustered BHs and subsequent rapid merging.
		\citet{Bellovary2019} find a similar phenomenon in zoom simulations of dwarf galaxies.
		Bellovary et al.~define ``overmerging'' to occur if either of the merging BHs were seeded $< 100$ Myr prior to the merger event.
		They suggest this time frame is long enough for BH feedback from existing BHs to suppress future BH formation.
		We find approximately $35\%$ of overmassive BHs and $15\%$ of median BHs have experienced an overmerging event.
        Since BHs that undergo BH mergers have a subsequently higher accretion rate, such overmerging may unphysically contribute to the BH mass.
        However, we find overmerging does not guarantee that a BH will become overmassive or grow to high $M\textsubscript{BH}$.

		Finally, current observations of BHs in dwarf galaxies indicate BHs may have masses lower than the BH seed mass used in this work~\citep{Reines2015,Baldassare2015,Nguyen2017,Nguyen2018,Nguyen2020,Schutte2019}. Observed BH masses in $M\textsubscript{star} \sim 10^{9}$ galaxies are typically a few $\times 10^5 M_\odot$, but can reach as low as $6.8 \times 10^{3} M_\odot$~\citep{Nguyen2020}.
		\citet{Mezcua2018} combine Chandra data of $z \lesssim 2.4$ dwarf galaxies with the $M\textsubscript{BH} - M\textsubscript{star}$ relation from \citet{Reines2015} and calculate BH masses consistent with the undermassive but not the overmassive BH masses in this work.
		In particular, overmassive BHs in \Rom{} may be unrealistically massive and accrete too much, causing us to overestimate their ability to drive galaxy-scale changes through feedback.
		Moving to higher resolution simulations in the future may give a more clear understanding of both how BHs grow within dwarf galaxies, and how dwarf galaxies may have their star formation suppressed by the BH.

\section{Summary} \label{sec:summary}
	We explore the connections between BH growth and dwarf galaxy evolution within the \Rom{} cosmological simulation.
	We investigate the source of scatter in the $M\textsubscript{BH} - M\textsubscript{star}$ relation and classify BHs as overmassive, undermassive, or median for their host $M\textsubscript{star}$.
	Using these classifications, we follow the primary growth modes for both BHs and their hosts.
	We can summarize our results as follows:

	\begin{itemize}
		\item
			\Rom{} forms massive BHs at early times in well-resolved dwarf galaxies ($10^{8} < M\textsubscript{star} < 10^{10} M_\odot$) that are consistent with observed BH scaling relations above $M\textsubscript{star} \gtrsim 10^{8.5} M_\odot$.
			The $M\textsubscript{BH} - M\textsubscript{star}$ relation shows a high degree of scatter in galaxies below $M\textsubscript{star} < 10^{10} M_\odot$.

		\item
			The scatter in the $M\textsubscript{BH} - M\textsubscript{star}$ relation is tied to the BH primary growth mode and likely to the initial growth environment.
			BHs that end up in the bottom quartile in $M\textsubscript{BH}$ by $z=0.05$ (undermassive BHs) have experienced almost no BH mergers, and instead grow primarily through low accretion rates.
			BHs that end up in the top quartile (overmassive BHs) experience at least one BH merger and undergo more accretion.
			Although overmassive BHs accrete more than their undermassive BH counterparts, BHs in dwarf galaxies grow little relative to those found in massive galaxies.

		\item
			The efficiency of BH growth within dwarf galaxies depends on the host formation history, though the difference is most pronounced in galaxies with $M\textsubscript{star} > 10^{9} M_\odot$.
			Hosts of overmassive BHs rapidly build up dark matter and stars, and experience suppression of star formation in their central regions around $z=2$.
			By $z=0.05$, hosts of overmassive BHs have a lower central density of stars than their undermassive BH counterparts.
			Undermassive BH hosts and galaxies without BHs build up their stars and dark matter nearly identically, suggesting undermassive BHs do not significantly alter the properties of their host galaxies.

		\item
			Above $M\textsubscript{star} > 10^{9} M_\odot$, quenching of star formation only occurs in galaxies that host BHs, and the majority of such galaxies host an overmassive BH\@.
			Overmassive BH hosts show significantly lower levels of \HI gas content, regardless of stellar mass, relative to undermassive BH hosts.
			Further, hosts of overmassive BHs exhibit higher fractions of BH to SNe energy injection than undermassive BH hosts, suggesting overmassive BHs have significant impact on the evolution of dwarf galaxies.
	\end{itemize}

		A substantial fraction ($\sim40$\%) of the BHs in our low-mass galaxy sample grow via mergers with other BHs and exhibit little total growth by accretion.
		Overall, our results depict a view of BH seeds forming in low-mass galaxies which do not foster efficient gas accretion very frequently.
		Consequently, the most efficient way to grow BHs in many small galaxies is through mergers with other BHs.
		Once a galaxy becomes large enough to have a deeper potential well, BH growth by gas accretion may happen more efficiently.
		Tests of the multiple early BH mergers found in \Rom{} will be done in the future by the Laser Interferometric Space Antenna, which will detect BH-BH mergers with total masses $10^4-10^7 M_\odot$ up to $z \sim 20$ with good signal-to-noise~\citep{Amaro-Seoane2017}.

\section{Acknowledgements}
	We would like to thank the referee for their thorough and insightful comments.
	Ray Sharma thanks Ena Choi, Jenny Greene, and Amy Reines for helpful discussions.
	This material is based upon work supported by the National Science Foundation under Grant No. NSF-AST-1813871.
	JMB is grateful for support from NSF grant AST-1812642.
	\Rom{} is part of the Blue Waters sustained-petascale computing project, which is supported by the National Science Foundation (awards OCI-0725070 and ACI-1238993) and the state of Illinois.
	Blue Waters is a joint effort of the University of Illinois at Urbana–Champaign and its National Center for Supercomputing Applications.

\bibstyle{apj}
\bibliography{library}
\end{document}